\documentclass[5p]{elsarticle}

\usepackage{amssymb}
\usepackage{amsmath}
\usepackage{slashbox}

\usepackage{bm}
\usepackage{graphicx}

\begin{document}

\title{Physical Phenomenology of Phyllotaxis}

\author{Takuya Okabe}
\address{
Faculty of Engineering, Shizuoka University, 
3-5-1 Johoku, 
Hamamatsu 432-8561,Japan}

\begin{abstract}

We propose an evolutionary mechanism of phyllotaxis,  regular arrangement of
 leaves on a plant stem. 
It is shown that 
the phyllotactic pattern with the Fibonacci sequence has a selective
 advantage, 
for it involves 
the least number of phyllotactic transitions
during plant growth.

\end{abstract}

\begin{keyword}
Schimper-Braun; Fibonacci number; Stern-Brocot tree; natural selection
\end{keyword}

\maketitle

\section{Introduction}

\begin{figure}
\begin{center}
\includegraphics[width=.4 \textwidth]{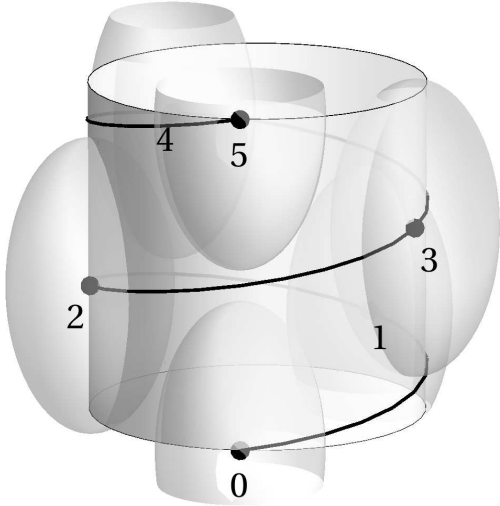}
\caption{\label{para25}
A pattern with a phyllotactic fraction (PF) $\alpha=2/5$. 
The angle between consecutive leaves (points) is 2/5 of a rotation.
As a guide to the eye,
a spiral, called the ontogenetic or genetic spiral, is drawn 
through the points in their order of succession. 
Repulsive interaction between the points is represented by 
(imaginary) prolate spheroids around the points.
}
\end{center}
\end{figure}

Phyllotaxis, regular arrangement of leaves on a plant stem, 
has since long attracted the minds of botanists, mathematicians and
physicists (\cite{dt17,jean94,abj97,kuhlemier07}).
Most commonly, alternate leaves along a twig 
execute 
a spiral with an angle of 1/2, 1/3, 2/5, 3/8
 of a full rotation, 
or otherwise with a limit angle of 137.5 degrees  (Fig.~\ref{para25}).
Patterns with the other fractions are also
observed, though uncommonly.
Hence
phyllotaxis is regarded not as a universal {\it law} but as a fascinatingly
prevalent {\it tendency} (\cite{coxeter61}).
For decades, mathematical works have elucidated 
number theoretical structure of phyllotaxis 
and deepened our understanding of the subject
significantly (\cite{coxeter72,adler74,ridley82,mk83,ropl84,rk89a,rk89b,levitov91b}).
Nevertheless, there still remains a fundamental problem of why and how
only some specific numbers are favored by nature.
This is a problem of natural science.

Physical or chemical models studied thus far are mostly based on 
{\it dynamical} mechanism, 
according to which a phyllotactic pattern is supposed to be realized
autonomously 
as an end result
of its own mechanical or chemical 
dynamics (\cite{adler74,thornley75,rk89b,levitov91b,dc96}). 
This is mechanical determinism.
Mathematicians and physicists generally appreciate this point of view. 
On the contrary, biologists generally take the opposite viewpoint that
there are so strong reasons for the plants to have genetic information
that a {divergence angle} between adjacent leaves must be determined genetically.
This is genetic determinism.
In this case, we still have to make clear 
how 
the plants inherit the mathematical characteristics, i.e., 
a {\it static} or statistic mechanism of phyllotaxis is asked for. 
When it comes down to it, 
 there seems no hope to do without mathematics, 
for we have to explain why the divergence angle is 
programmed to take neither
130$^\circ$ nor 140$^\circ$  but the magic number
137.5$^\circ$. 
%
%
%
Unfortunately, 
candidate explanations 
have been only descriptive and qualitative, 
%
or otherwise
quantitative but {\it teleological} such that 
packing efficiency (\cite{ridley82}) 
or uniformity (\cite{mk83,ropl84,bryntsev04})  is meant to be maximized
resultingly.  
In any case, 
at this fundamental level, 
we have to regard phyllotaxis 
as a conundrum of theoretical physics (or theoretical biology),
for we have to deal with mathematics observed in the real world.
The aim of this paper is to present a satisfactory static mechanism
 to be contrasted not only with the 
existing dynamical models but  with the
teleological explanations. 



We propose a new mechanism 
 based on 
a growth model 
of a plant.
The model is basically characterized by two parameters, 
an initial (preset) value of the {divergence angle}, $2\pi \alpha_{0}$, 
and a vertical range of repulsive interaction, $n_{c}$.
The model is based on the observation that 
a plant has inherent abilities (1) to arrange leaves primarily
on a regular spiral 
with a constant angle of rotation $2\pi \alpha_{0}$, 
and (2) to exert secondary torsions 
between the leaves within reach of $n_{c}$.
The former is consistent with experiments on apical meristems
(\cite{fujita39,snow62}). 
The latter is supposed to operate in a vascular system (\cite{larson77}).
%
%
The secondary interaction reinforces the regularity of the helical
arrangement, 
thereby the observed angle $\alpha$ generally undergoes changes from 
the initial value $\alpha_0$. 
With the vertical range $n_{c}$ regarded as a growth index, 
it is shown that $\alpha$ goes through stepwise transitions 
between phyllotactic fractions (PF)
in the course of the growth (Fig.~\ref{paratransition}). 
In fact, many plants progress through distinct 
phyllotactic transition (PT) during early stages of development.
It should be rather remarked that 
the secondary changes,  essential for us, 
are usually noticed but often disregarded as irrelevant (as secondary).
%
To outline the proposed mechanism, 
consider a population of random samples with 
all possible values of the preset angle $\alpha_{0}$, 
and let them grow according to the model. 
As they grow,  each sample will exhibit its own PTs
respectively, depending on $\alpha_{0}$.
With all the grown samples, 
we leave from formal analysis. 
At this point, 
we appeal to biological reasoning
according to 
{natural selection}, by
which 
all but that with the golden angle $2\pi\alpha_0 =137.5^\circ$ turns out
to be eliminated.
The selected sample is favored in nature, 
for it is structurally the most stable because it undergoes the least PTs. 
To put it concretely, 
we show how the number of PTs, $N_{\rm PT}$, 
depends on the preset angle $\alpha_{0}$ and the growth index $n_{c}$ (Fig.~\ref{phyllo}). 
To the author's knowledge, 
no such mathematical {\it evolutionary} mechanism of phyllotaxis has ever been put forward. 





\section{Model}
\label{sec:model}


\begin{figure}
\begin{center}
\includegraphics[width=0.3\textwidth]{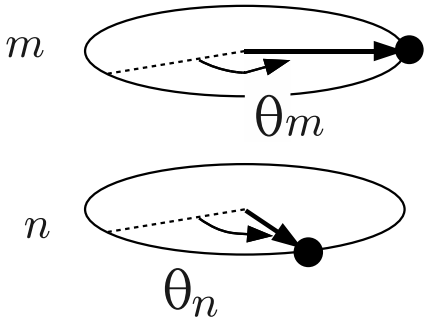}
\caption{\label{vn}
Interaction $V_n(\theta_m-\theta_n, m-n)$ 
between the points $(m,\theta_m)$ and $(n,\theta_n)$ is 
periodic in $\theta_m-\theta_n$ with a period of $2\pi$. 
By definition of $X$ and $n_{c}$, 
$V_n(\theta_m-\theta_n, m-n)$ 
vanishes for $X< |\theta_m-\theta_n|/2\pi < 1 $
and $|m-n|> n_{c}$.
}
\end{center}
\end{figure}

We restrict ourselves to the most common case of a spiral or helical pattern with 
a single `leaf' at each level (node).
We use an integer $n$ to label successive points
along a genetic spiral (Fig.~\ref{para25}), 
whose positions are given by the coordinates $(\theta_n, n)$ 
or by $(x_n, n)$ in terms of $\theta_n=2\pi x_n$. 
In this dimensionless representation
with normalized length scales, 
the coordinate may be regarded to represent the position of
either 
leaf primordia on apical meristem or 
leaf traces in vascular system. 
Here the important point is that physical quantities are 
periodic with respect to the angular coordinate $\theta_n$.
The angle $\theta_n$ is measured from $\theta_0=0$ for $n=0$, 
and is regarded to take a value within $-\pi \le \theta_n <\pi $ (or $-1/2 \le x_n< 1/2$). 
To describe phenomenologically a torsional force between two points $(\theta_n, n)$ and
$(\theta_m, m)$ $(n\ne m)$, 
we introduce repulsive interaction $V_n (\theta_m-\theta_n, m-n)$ (Fig.~\ref{vn}). 
As a theoretical and phenomenological tool to clarify number theoretical structure of the macroscopic phenomenon,   
details for implementation of the interaction need not be specified.


For the sake of simplicity and convenience, let us write
\begin{equation}
V_n (2\pi x , m) = u_n V  (2\pi x, m) =  u_n v_m V  (x).
\label{Vn2pixmunv2pixmunvmvx}
\end{equation}
The angular dependence is represented by $V (x)$ 
and the vertical (internode) dependence  is described with $v_m$, 
both of which are defined by the last equation.
The factor $u_n$ 
represents the dependence on the subscript $n$ of $V_n (2\pi x , m)$, 
which occurs 
because translational invariance along the stem 
(in the vertical direction) is broken in
general.
However, it turns out that the factor $u_n$ drops out of our problem.


The interaction $V_n (\theta,m)$ is characterized by two parameters. 
They are finite ranges of $V_n (\theta,m)$, 
in the angular direction
$\theta$ (or $x=\theta/(2\pi)$), and 
in the vertical direction $m$. 
For the former, we introduce a half-width $X$ $(<1/2)$ of $V(x)$, 
i.e., 
\begin{eqnarray}
 V(x) &>  & 0, \qquad   0\le |x|<  X, \nonumber\\
 V(x) &\simeq & 0, \qquad    X< |x| <1/2.
\nonumber
\end{eqnarray}
For example, let us use 
\begin{equation}
 V (x )  =  e^{-\left(\frac{2x}{X}\right)^2}, \qquad |x|\le 1/2.
\label{V(x)}
\end{equation}
Note that $V(x)$ is periodic so that $V(x)=V(x+1)$,
and we may set $V(0)=1$ arbitrarily.
In terms of the width $X$ thus defined,  the angular width
of the original interaction $V_n (\theta,m)$ 
is given by 
\begin{equation}
\Delta \theta \simeq 4\pi X.
\label{Deltathetasimeq4piX}
\end{equation}
The second parameter is the vertical range of influence 
$n_{c}$.
By definition, we have $V_n (\theta, m)>   0$ 
for $0< m \le n_{c}$  and 
$V_n (\theta,m)\simeq  0$  for $m > n_{c}$, 
or 
\begin{eqnarray}
 v_n &> & 0, \qquad   0< n \le n_{c}, 
\nonumber\\
 v_n  &\simeq & 0, \qquad    n_{c}< n. 
\nonumber
\end{eqnarray}
In what follows, $n_{c}$ plays an important role.




We investigate a regular spiral arrangement with the {\it divergence angle}
$2\pi \alpha$, namely, 
\begin{equation}
 \theta_n = 2\pi n \alpha. 
\qquad (x_n= n \alpha.)
\label{thetan=2pinalpha}
\end{equation}
In principle, 
the phyllotactic index 
$\alpha$ may take any real number  (Fig.~\ref{para25}).
Nevertheless, 
in our model, and in real life, $\alpha$ turns out to be a fraction, 
called {\it phyllotactic fraction} (PF).
Phenomenologically, 
$\alpha$ obeys 
a mechanical relaxation equation, 
\begin{equation}
\frac{d \alpha}{dt} =- \frac{d E}{d \alpha}.
\label{taudalphadt=-dedalpha}
\end{equation}
The right-hand side is a torsional force, 
represented by the total interaction $E$, 
\begin{eqnarray}
 E&=&\sum_{n=0}^\infty \sum_{m > n } V_n (\theta_m-\theta_n, m-n)
\label{E=sumn0}
\\
&=&\sum_{n=0}^\infty \sum_{m =1 }^\infty V_n (2\pi m\alpha, m).
\end{eqnarray}
Substituting Eq.~(\ref{Vn2pixmunv2pixmunvmvx}), 
we obtain
\begin{equation}
E=  v(\alpha) \sum_{n=0}^\infty u_n, 
\label{E=valpha=sumn0infty}
\end{equation}
where 
\begin{equation}
 v(\alpha) = \sum_{m =1 }^\infty  v_m V ( m\alpha).
\label{valphaequiv}
\end{equation}
The factor $\sum_{n=0}^\infty u_n$ in Eq.~(\ref{E=valpha=sumn0infty}) 
may be dropped hereafter, as it is a constant independent of $\alpha$.
Substituting Eq.~(\ref{E=valpha=sumn0infty}) in Eq.~(\ref{taudalphadt=-dedalpha}), 
we find the phyllotactic index $\alpha$ in static equilibrium as a minimum of
the effective interaction $v(\alpha)$.
For instance, 
when $v(\alpha)$ has a single minimum at $\bar{\alpha}$, 
we may assume a parabolic potential 
$E=\frac{k}{2}(\alpha-\bar{\alpha})^2$ 
around the minimum $\alpha\simeq \bar{\alpha}$. 
Then 
we obtain $\frac{d \alpha}{dt} =- k (\alpha-\bar{\alpha})$ from
Eq.~(\ref{taudalphadt=-dedalpha}),
and 
$\alpha(t)=\bar{\alpha} + (\alpha(0)-\bar{\alpha} ) e^{-kt}$ 
as a solution. 
Therefore, in the end, we reach 
static equilibrium at the minimum $\alpha(\infty)=\bar{\alpha}$, 
irrespective of the initial value $\alpha(0)$.
In general, $v(\alpha)$ may have many local minima. 
To which minimum $\alpha$ evolves into depends on the initial value $\alpha(0)$.
This is the third parameter $\alpha_{0}$, 
\begin{equation}
 \alpha(0)= \alpha_{0}.
\label{alpha0=alphaini}
\end{equation}

We introduced several quantities 
to define our model.
Among others, 
 $n_c$ and $ \alpha_{0}$ are the most important.
Main results given below are not affected essentially 
 by the other quantities as $X$ and $v_m$ for $v(\alpha)$ in
Eq.~(\ref{valphaequiv}).
In the next section, 
we investigate conditions to realize a local minimum of $v(\alpha)$.
%
Hereafter we restrict ourselves to 
$0<\alpha \le 1/2$, 
because $v(1-\alpha)=v(-\alpha) = v(\alpha)$
by bilateral symmetry.

\section{Results}
\label{sec:result}

\subsection{Phyllotactic Fraction (PF)}
\label{sec:pf}

\begin{figure}
\begin{center}
\includegraphics{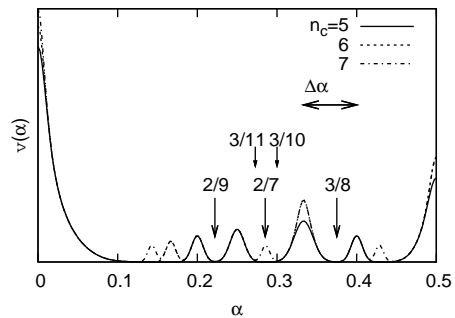}
\caption{\label{pot5-7}
Effective interaction 
$v(\alpha)$ 
for $n_{c}=5,6$ and 7
($X=0.1$ and $v_n=0.8^n$ for $n\le n_{c}$). 
To reach a minimum at $\alpha=3/8$, 
the initial value $\alpha_0$ must fall within $\Delta \alpha$ denoted by
 the double-headed arrow. 
}
\end{center}
\end{figure}
By way of illustration, 
$v(\alpha)$
in Eq.~(\ref{valphaequiv})
 is plotted 
for $n_{c}=5,6$ and 7 
in Fig.~\ref{pot5-7}, 
where we use 
Eq.~(\ref{V(x)}) with $X=0.1$ and $v_n=0.8^n$ for $n\le n_{c}$. 

Consider the case $n_{c}=5$,  the solid curve in Fig.~\ref{pot5-7}. 
We observe that $v(\alpha)$  has five minima. 
In effect, they lie around fractions $\alpha =1/6, 2/9, 2/7, 3/8$
and $3/7$.
Let us increase $n_{c}$ from 5 to 7 to see if the minima are affected.
For $n_{c}=7$, 
the two minima at 2/9 and 3/8  remain almost intact, 
whereas the other three minima 1/6, 2/7 and 3/7 
for $n_c=5$ are lost (Fig.~\ref{pot5-7}).

To reach the minimum at PF 3/8, 
the initial value $\alpha_{0}$ has to be in a range  $1/3\lesssim \alpha_{0}\lesssim 2/5$.
Let us introduce the total width $\Delta \alpha$ of a range allowed for
$\alpha_{0}$.
For PF 3/8, 
we have $\Delta \alpha = \frac{2}{5}-\frac{1}{3}\simeq 0.067$. 
The range is indicated with the double headed arrow in Fig.~\ref{pot5-7}.
%
Similarly, we get $\Delta \alpha\simeq 0.05$ for 2/9, 
which is narrower than $\Delta \alpha \simeq 0.067$ for 3/8. 
Therefore, if $\alpha_{0}$ is to be chosen randomly for a fixed $n_{c}$, 
it would be (0.067/0.05=1.3 times) easier to realize PF 3/8 than to
realize 2/9.


From Fig.~\ref{pot5-7}, we note that 
the minimum at 2/7 for $n_{c}=5$ and 6
turns into a local {\it maximum} for $n_{c}=7$. 
Similarly, 
the minimum at $\alpha=3/8$ 
becomes a {maximum} for $n_{c}=8$ (not shown in Fig.~\ref{pot5-7}), 
whereas the minimum at 2/9 remains intact 
for $n_{c}=5,6,7$ and 8. 
For each PF, 
we define $n_{0}$ and $\Delta n$ by the condition 
\begin{equation}
n_{0} \le n_{c}\le n_{0}+\Delta n
\label{nmax0lenmaxlenmax0+deltanmax}
\end{equation}
for the fraction to be a minimum.
We obtain $(n_{0}, \Delta n)=(5,2)$ for 3/8, 
and $(n_{0}, \Delta n)=(5,3)$ for 2/9.

When a minimum becomes a maximum, 
two new minima appear on both sides of the maximum.
For example, 
the local minimum at  $\alpha =2/7$ for $n_{c}=5,6$ ($(n_{0}, \Delta n)=(4,2)$)
becomes a local maximum at $n_{c}=7$. 
When 
a `mother' PF 2/7 is lost to become the maximum, 
it is flanked on both sides 
by two newborn `daughter' minima at 3/11 and 3/10 (Fig.~\ref{pot5-7}).
In other words, 
the mother PF $\alpha=2/7$ branches (fissions) into 
the daughter PFs 3/11 and 3/10 at $n_{c}=7$.

\begin{figure}
\begin{center}
\includegraphics{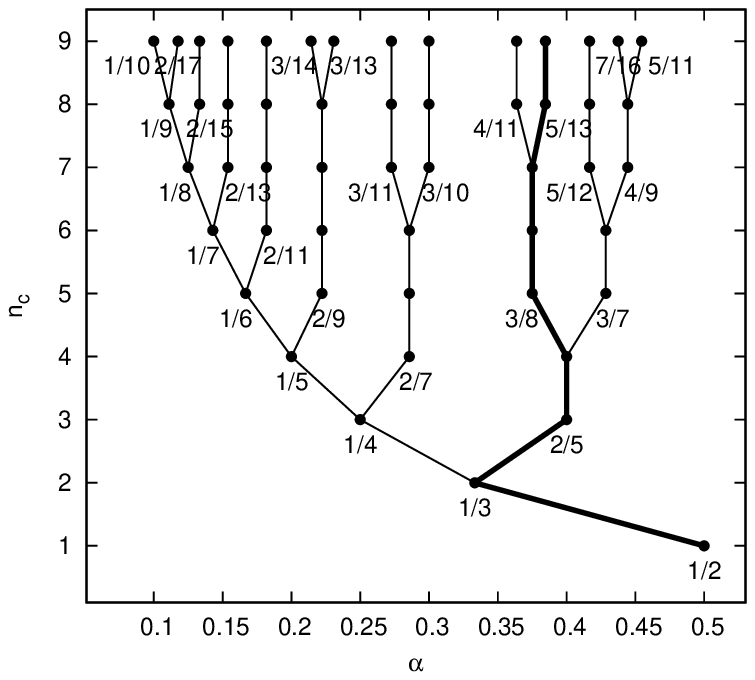}
\caption{\label{tree}
Position $\alpha$ of a local minimum of $v(\alpha)$ 
is represented by a point
labeled with a fraction. 
Lines connecting the points 
represent hierarchical branching structure of the fractions. 
The length of the vertical segment 
stretching from a point
represents $\Delta n$ for the fraction
(e.g., $\Delta n=0$ for 1/4 and $\Delta n=1$ for 2/5). 
Fractions on the bold line comprise the main sequence of phyllotaxis. 
}
\end{center}
\end{figure}
\begin{figure}
\begin{center}
\includegraphics[width=0.48\textwidth]{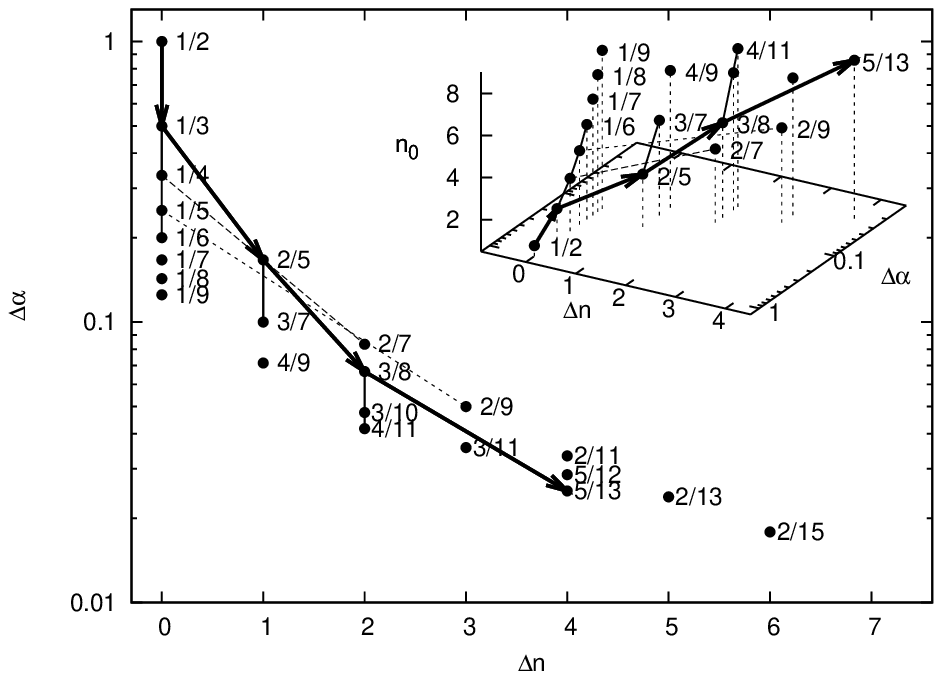}
\caption{\label{deln-alini}
For lower order fractions in Fig.~\ref{tree}, 
$\Delta \alpha$ is plotted against  $\Delta n$.
The inset shows a 3D plot of  $n_{0}$ against 
$\Delta n$ and $\Delta \alpha$
(defined in Sec.~\ref{sec:pf}). 
The bold arrows represent the main sequence. 
}
\end{center}
\end{figure}

In this manner, we can locate the positions of the local minima 
of $v(\alpha)$ 
and construct their branching structure.
They are shown 
in Fig.~\ref{tree}, 
where each dot representing a minimum is labeled with a fraction (PF).
In Fig.~\ref{tree}, 
there is a vertical segment stretching upward from each dot.
The length of the segment is $\Delta n$ for the fraction.
In Fig.~\ref{deln-alini}, we plot 
the fractions in the $\Delta n$-$\Delta \alpha$ plane, 
as well as in the $\Delta n$-$\Delta \alpha$-$n_0$ space.
In the figures, the observed main sequence of phyllotaxis is drawn with the bold line. 


In the above and the following, 
we make good use of the interesting and important property of the model that 
the divergence angle
$\alpha$ of a phyllotactic pattern 
is given by a fraction (PF). 
Indeed, deviation from an exact fraction may be expected in
general. 
This is discussed in Sec.~\ref{sec:optimalwidth}.


\subsection{Phyllotactic Transition (PT)}

\begin{figure}
\begin{center}
\includegraphics{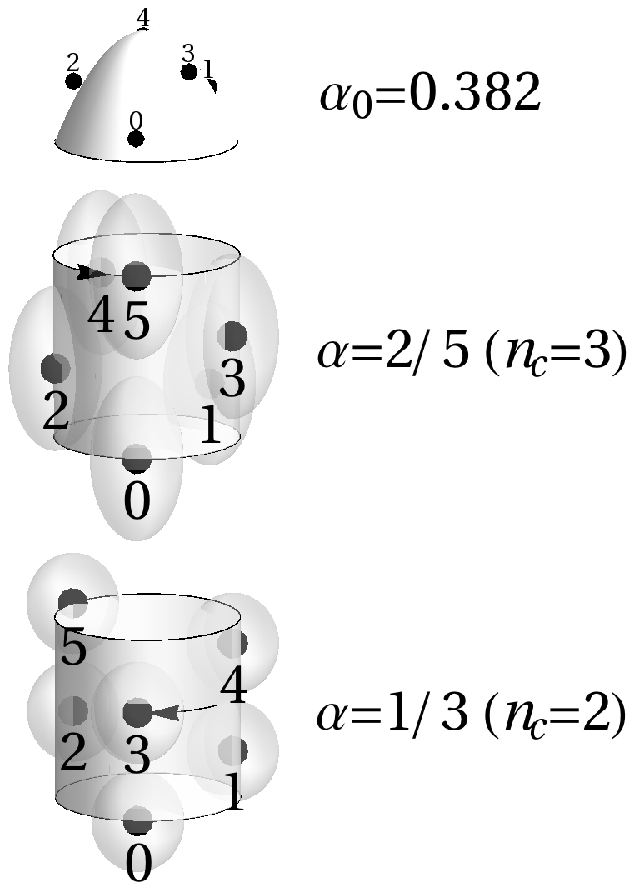}
\caption{\label{paratransition}
An initial pattern ($\alpha_0=0.382$) of leaf primordia on the apex (top). 
A `developing' leaf pattern for $\alpha=2/5$ ($n_c=3$) (middle)
lies on top of a `mature' leaf pattern for $\alpha=1/3$ ($n_c=2$) (bottom). 
Torsional motion is indicated by an arrow.
}
\end{center}
\end{figure}

To compare with a real plant, 
we regard $n_{c}$ as a growth index.
This is reasonable because 
the increase of $n_{c}$ in our normalized (cylinder) representation
corresponds to a decrease of plastochron ratio in a disc
representation (\cite{richards51}), 
and to a decrease of the internode distance 
if it were introduced explicitly (\cite{adler74}).
The index $n_c$, presumably related to the length of leaf-trace
primordia,  is supposed to increase in the course of plant growth
so that a stem is vertically organized into zones with different values of $n_c$. 
In a developing leaf zone near the apex, $n_c$ is large.
In a mature leaf zone near the plant base,  $n_c$ is small. 
%
%
For a given value of $\alpha_0$,  
we obtain patterns for different $n_c$ standing in a row along the
stem, as  illustrated in Fig.~\ref{paratransition}.
Thus we explain {\it phyllotactic transition} (PT), 
the transition of $\alpha$ between different PFs along the stem.


\begin{figure}
\begin{center}
\includegraphics[width=.5 \textwidth]{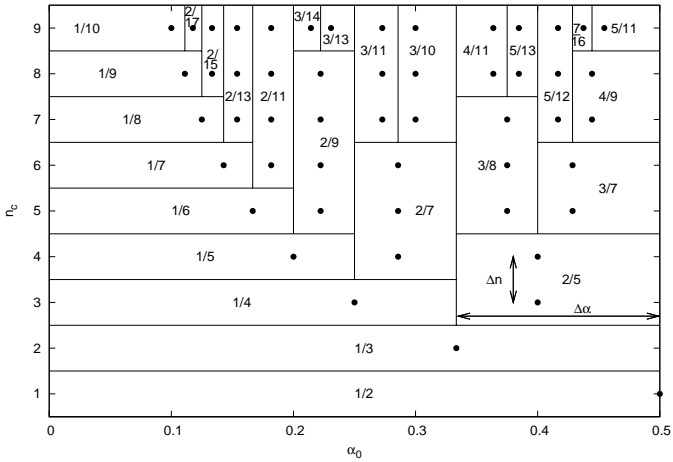}
\caption{\label{treePF}
Diagram for phyllotactic fractions (PF). 
PF $\alpha=2/5$ is obtained for $n_c=3,4$ and $1/3< \alpha_0<1/2$
$(\Delta n=1, \Delta \alpha=0.17)$. 
}
\end{center}
\end{figure}
For given $n_{c}$, from Fig.~\ref{tree}
we can read $\alpha$ resulting from arbitrary $\alpha_{0}$. 
Firstly, locate the reference point $(\alpha_0, n_{c})$ in the figure.
Then, the PF $\alpha$ is selected 
from the neighboring two minima on both sides of the reference point
by comparing the horizontal coordinate $\alpha_0$ with 
that of a maximum between the two minima.
If $\alpha_0$ is on the left (right) of the maximum,  
then the minimum in the left (right) hand side is chosen.
For example, for $\alpha_0\simeq 0.3$, 
we get 
$\alpha=1/2$ for $n_{c}=1$, and 
$\alpha=1/3$ for $n_{c}=2$. 
For $n_c=3$, 
we choose $\alpha=1/4$ from the two minima 1/4 and 2/5 because 
$\alpha_0\simeq 0.3< 1/3$, the maximum between the two minima.

\begin{figure}
\begin{center}
\includegraphics[width=.58\textwidth]{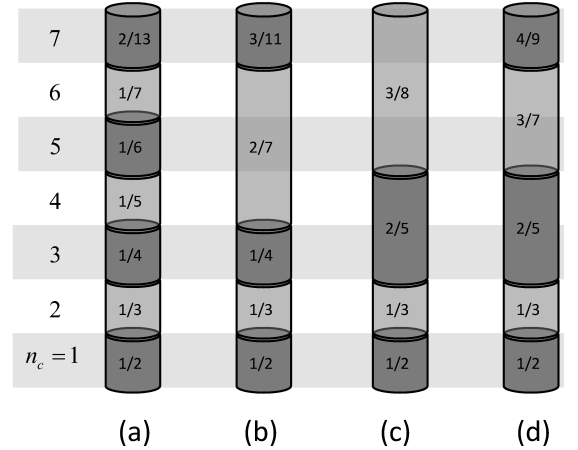}
\caption{\label{phyllo}
Lower part of 
four sequences in Table~\ref{alphanmax} are displayed vertically. 
(a) $\alpha_{0}=0.146317$ ($N_{\rm PT}=6$), 
(b) $\alpha_{0}=0.281978$ ($N_{\rm PT}=4$), 
(c) $\alpha_{0}=0.375791$ ($N_{\rm PT}=3$), 
(d) $\alpha_{0}=0.469168$ ($N_{\rm PT}=4$).
The most favorable is (c) with the least number of phyllotactic
 transitions ($N_{\rm PT}=3$).
}
\end{center}
\end{figure}

As a result, we obtain a diagram in Fig.~\ref{treePF}.    
The diagram may serve to get PF $\alpha$ for arbitrary pair $(\alpha_0, n_c)$.
In Table~\ref{alphanmax}, sequences of PF $\alpha$ 
for randomly chosen values of $\alpha_{0}$
are given representatively. 
With real plants in mind, 
four of them 
are displayed in Fig.~\ref{phyllo}.
\begin{table*}[t]
\begin{center}
\caption{\label{alphanmax} 
Sequences of PF $\alpha$ resulting from 
arbitrarily chosen initial values of $\alpha_{0}$.
$N_{\rm PT}$ in the last column 
 counts the number of phyllotactic transitions (the number
 of times PF $\alpha$ changes) for $1\le n_{c} \le 17$.
}
\begin{tabular}{|c|ccccccccccccc|c|}\hline
\backslashbox{$\alpha_{0}$}{$n_{c}$} 
& 1 & 2 &3&4&5&6&7&8&9&10&11 &13& 17 &$N_{\rm PT}$ \\ \hline
0.025714 & $\frac{1}{2}$ &
	 $\frac{1}{3}$&$\frac{1}{4}$&$\frac{1}{5}$&$\frac{1}{6}$&$\frac{1}{7}$&$\frac{1}{8}$&$\frac{1}{9}$&$\frac{1}{10}$&$\frac{1}{11}$&$\frac{1}{12}$
					    						&$\frac{1}{14}$&$\frac{1}{18}\strut$&16\\ \hline
0.104117  & $\frac{1}{2}$ & $\frac{1}{3}$&$\frac{1}{4}$&$\frac{1}{5}$&$\frac{1}{6}$&$\frac{1}{7}$&$\frac{1}{8}$&$\frac{1}{9}$&$\frac{1}{10}$&$\frac{2}{19}$ &$\frac{2}{19}$&$\frac{2}{19}$&$\frac{2}{19}\strut$&9 \\ \hline
0.146317 & $\frac{1}{2}$ & $\frac{1}{3}$&$\frac{1}{4}$&$\frac{1}{5}$&$\frac{1}{6}$&$\frac{1}{7}$& $\frac{2}{13}$ &$\frac{2}{13}$ &$\frac{2}{13}$ &$\frac{2}{13}$ &$\frac{2}{13}$&$\frac{3}{20}$& $\frac{3}{20}$&7 \\ \hline
0.175101& $\frac{1}{2}$ & $\frac{1}{3}$&$\frac{1}{4}$&$\frac{1}{5}$&$\frac{1}{6}$& $\frac{2}{11}$ &$\frac{2}{11}$ &$\frac{2}{11}$ &$\frac{2}{11}$ &$\frac{2}{11}$ &$\frac{3}{17}$ &$\frac{3}{17}$& $\frac{4}{23}$&7\\ \hline
0.281978 & $\frac{1}{2}$ & $\frac{1}{3}$&$\frac{1}{4}$& $\frac{2}{7}$ &$\frac{2}{7}$&$\frac{2}{7}$ &$\frac{3}{11}$ &$\frac{3}{11}$ &$\frac{3}{11}$ &$\frac{3}{11}$  & $\frac{5}{18}$&$\frac{5}{18}$&$\frac{5}{18}$& 5 \\ \hline
0.286878  & $\frac{1}{2}$ & $\frac{1}{3}$&$\frac{1}{4}$& $\frac{2}{7}$ &$\frac{2}{7}$&$\frac{2}{7}$ &$\frac{3}{10}$ &$\frac{3}{10}$ &$\frac{3}{10}$ &$\frac{5}{17}$  & $\frac{5}{17}$ &$\frac{5}{17}$&$\frac{7}{24}$&6 \\ \hline
0.305352  & $\frac{1}{2}$ & $\frac{1}{3}$&$\frac{1}{4}$& $\frac{2}{7}$ &$\frac{2}{7}$ &$\frac{2}{7}$ &$\frac{3}{10}$ &$\frac{3}{10}$ &$\frac{3}{10}$ &$\frac{4}{13}$ &$\frac{4}{13}$  &$\frac{7}{23}$&$\frac{7}{23}$&6 \\ \hline
0.375791  & $\frac{1}{2}$ & $\frac{1}{3}$&$\frac{2}{5}$& $\frac{2}{5}$ &$\frac{3}{8}$ &$\frac{3}{8}$ &$\frac{3}{8}$ & $\frac{5}{13}$ &$\frac{5}{13}$ &$\frac{5}{13}$  & $\frac{5}{13}$    &$\frac{8}{21}$ &$\frac{8}{21}$ &5  \\ \hline
0.437801  & $\frac{1}{2}$ & $\frac{1}{3}$&$\frac{2}{5}$& $\frac{2}{5}$ &$\frac{3}{7}$ &$\frac{3}{7}$ &$\frac{4}{9}$ & $\frac{4}{9}$ &$\frac{7}{16}$ &$\frac{7}{16}$ &$\frac{7}{16}$  &$\frac{7}{16}$ &$\frac{11}{25}$& 6  \\ \hline
0.469168  & $\frac{1}{2}$ & $\frac{1}{3}$&$\frac{2}{5}$& $\frac{2}{5}$ &$\frac{3}{7}$ &$\frac{3}{7}$ &$\frac{4}{9}$ & $\frac{4}{9}$ &$\frac{5}{11}$ &$\frac{5}{11}$  &$\frac{6}{13}$   &$\frac{7}{15}$&$\frac{15}{32}$& 9 \\ \hline
\end{tabular}
\end{center}
\end{table*}


\subsection{Mechanism of Phyllotaxis}
\label{subsec:main}


Nature's preference of particular PFs is accounted for simply by the following
hypothesis:  
(H) {\it A phyllotactic pattern of a fraction with larger $\Delta n$ is more favored.}

%

%

For $n_{c}=3$ in Fig.~\ref{tree},  
we have two possible patterns $\alpha=1/4$ or 2/5,  
 depending on $\alpha_0$. 
According to (H), the latter 2/5 is favored, 
because  $\Delta n=1$ for 2/5
is larger than $\Delta n=0$ for 1/4.
In this manner, as $n_c$ is increased,  
the hypothesis  (H)  gives us a sequence 
on the thick line in Figs.~\ref{tree} and \ref{deln-alini}, 
that is, 
\begin{equation}
1/2, 1/3, 2/5, 3/8, 5/13, \cdots.
\label{fibonacchi0}
\end{equation}
This is the {\it main sequence} of phyllotaxis almost always observed. 
In fact, the main sequence covers more than 90\% of all observed cases (\cite{jean94}).
The numerator and the denominator of 
the PF in (\ref{fibonacchi0})
are alternate numbers of a Fibonacci sequence, 
\begin{equation}
1,1,2,3,5,8,13,21,34,55,89,144, \cdots, 
\label{fibonacchi}
\end{equation}
in which every number is the sum of the preceding two.
As a limit number of the sequence in (\ref{fibonacchi0}),  
we obtain 
$\alpha =(3-\sqrt{5})/2 \simeq 0.381966 $
($2\pi \alpha \simeq 137.5^\circ$), 
the golden angle or the Fibonacci angle. 

%




We argue that the hypothesis (H) is biologically plausible,  
because the larger $\Delta n$ ensures the more stability against  
expected variations of the growth index $n_{c}$. 
To show this explicitly, 
let us define $N_{\rm PT}$ 
as the number of PTs
encountered in a sequence for a given $\alpha_0$. 
$N_{\rm PT}$ counts how many times PF $\alpha$ changes as the index
$n_{c}$ is increased, 
so that it depends on $\alpha_{0}$ and the upper bound of $n_{c}$. 
Hence $N_{\rm PT}$ may be used as a measure of stability of 
a pattern with a given phyllotactic sequence. 
By definition, 
$N_{\rm PT}$ is small for a sequence comprised of PFs with small $\Delta n$.
Therefore,  (H) may be rephrased as follows: (H') {\it A favorable sequence has  small $N_{\rm PT}$.}

Biological implications of (H') may be understood intuitively from Fig.~\ref{phyllo}.  
According to (H'), 
the case (c) for $\alpha_{0}\simeq 0.38$ ($2\pi \alpha_{0}=137.5^\circ$) 
is the most favorable, because $N_{\rm PT}=3$ for (c) is the smallest of all. 
What this means must be quite obvious from the figure.  
We give $N_{\rm PT}$ in the last column of Table~\ref{alphanmax}. 
See the eighth row for $\alpha_0=0.375791$ 
in Table~\ref{alphanmax}.
We find that $N_{\rm PT}$  is the smallest 
for the main sequence, (\ref{fibonacchi0}),  for $\alpha_{0}\simeq 0.38$
($2\pi \alpha_{0}=137.5^\circ$).
As a second sequence with small $N_{\rm PT}$, 
we notice 
a sequence for $\alpha_{0}\simeq 0.28$ ($2\pi \alpha_{0}=99.5^\circ$), 
the fifth row in Table~\ref{alphanmax}, and (b) in Fig.~\ref{phyllo}. 
This sequence is observed but less commonly, and 
sometimes called the first accessory sequence.
In general, 
the main sequence always sets the lower limit of $N_{\rm PT}$, 
although there may be other sequences with the same lowest value.

\begin{figure}
\begin{center}
\includegraphics{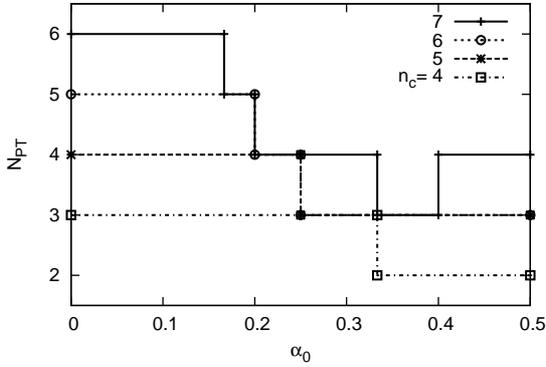}
\caption{\label{sb4-7}
The number of phyllotactic transitions $N_{\rm PT}$
for $n_{c} = 4$, 5, 6, and 7
are drawn against $\alpha_{0}$.
As $n_{c}$ increases to 7, 
initial values $\alpha_{0}$ 
within $1/3 < \alpha_{0}< 2/5$
($\Delta \alpha=0.07$)
are specifically selected because $N_{\rm PT}$ is the smallest there.
}
\end{center}
\end{figure}
\begin{figure}
\begin{center}
\includegraphics[width=.5\textwidth]{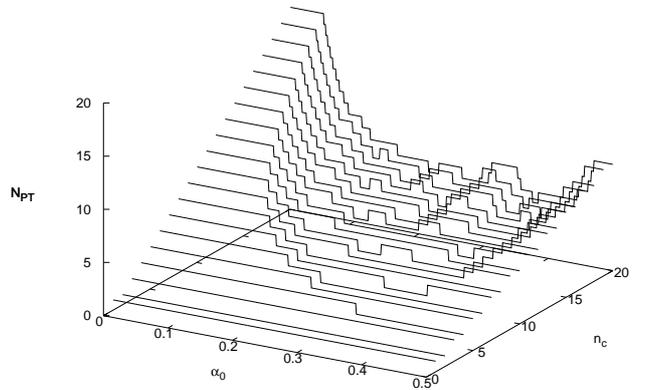}
\caption{\label{ancNpt}
$N_{\rm PT}$ as a function of $\alpha_0$ 
for $1\le n_c\le 20$ (cf. Fig.~\ref{sb4-7}).  
A narrow trough develops around $\alpha_{0} \simeq 0.38$
 ($2\pi\alpha_0 =137.5^\circ$), 
as $n_c$ increases.
}
\end{center}
\end{figure}

In Fig.~\ref{sb4-7}, $N_{\rm PT}$ for $n_{c}$ up to 4, 5, 6, and 7
is plotted against $\alpha_{0}$. 
The figure indicates how the samples with different $\alpha_{0}$
are discriminated as they grow. 
In the process of increasing $n_c$ to 4,  
any sample for $1/3< \alpha_0 <1/2$ $(\Delta \alpha=0.17)$ is more favorable 
than that for $0< \alpha_0 <1/3$, 
because the former has a smaller value $N_{\rm PT}=2$ 
than the latter with $N_{\rm PT}=3$. 
As we increase $n_c$ to 7 (the solid line in Fig.~\ref{sb4-7}),   
only restricted samples within a narrow window $1/3< \alpha_0 <2/5$
$(\Delta \alpha=0.07)$ 
are favored because of $N_{\rm PT}=3$,  which is the lowest value for $n_c=7$. 
In this filtering process, 
the width $\Delta \alpha$ for $\alpha_0$ decreases rapidly as $n_c$  increases.
In Fig.~\ref{ancNpt}, 
we plot $N_{\rm PT}$ against $\alpha_0$ and $n_c$.
The figure  indicates that
the minimum of $N_{\rm PT}$ with a decreasing width $\Delta \alpha$
develops around $\alpha_{0} \simeq 0.38$ ($2\pi\alpha_0 =137.5^\circ$)
as $n_c$ increases. 
In the end,  a single value is selected for $\alpha_0$, 
namely, 
the golden angle 
$\alpha_{0} =(3-\sqrt{5})/2 \simeq 0.381966 $ 
$(\Delta \alpha\rightarrow 0)$.

\begin{figure}
\begin{center}
\includegraphics{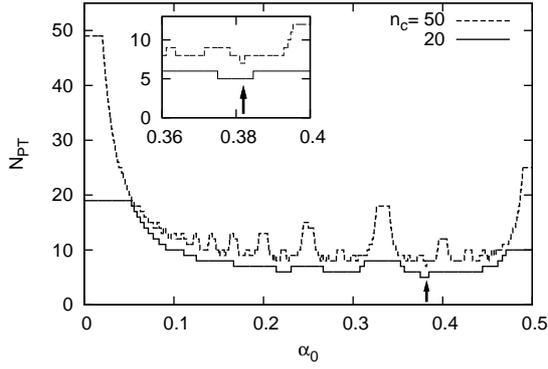}
\caption{\label{sb}
$N_{\rm PT}$  
for 
$n_{c} = 20$ (solid line) and 50 (dashed line).
The arrow at the minimum of  $N_{\rm PT}$
 indicates 
the golden angle
$\alpha_0 \simeq 0.3820$ ($2\pi\alpha=137.5^\circ$). 
}
\end{center}
\end{figure}
\begin{figure}
\begin{center}
\includegraphics{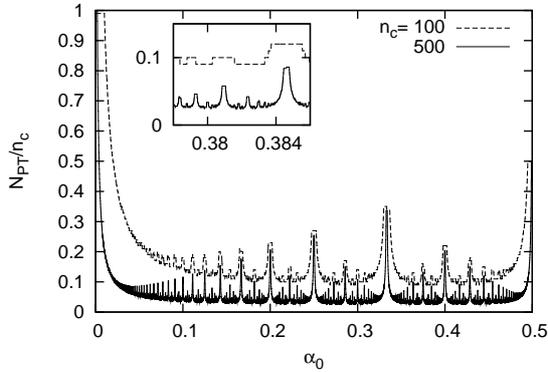}
\caption{\label{sb2}
$N_{\rm PT}/n_{c}$  for 
$n_{c} = 100$ (dashed line) and 500 (solid line).
The minimum at $\alpha_0 \simeq 0.38197$ 
is indiscernible 
for $n_{c}=500$ ($\Delta \alpha\simeq 10^{-5}$).
}
\end{center}
\end{figure}

In Fig.~\ref{sb}, we plot $N_{\rm PT}$ for $n_{c}$ up to 20 and 50, 
indicating how the stepwise ridges and troughs of $N_{\rm PT}$ develop as $n_{c}$ increases.
As $n_{c}$ increases, 
$\alpha_{0}$ should be gradually trapped in a minimum of $N_{\rm PT}$.
We find a small and shallow local minimum around $\alpha_{0} \simeq 0.22$
($2\pi\alpha_0 =78^\circ$), 
besides the absolute minimum at $\alpha_{0} \simeq 0.38$ and
the first accessory $\alpha_{0} \simeq 0.28$ mentioned above. 
These three cases are collectively called {normal phyllotaxis}
(Sec.~\ref{sec:majorseq} and (B.13)).
It is  remarked that $N_{\rm PT}$ (or $N_{\rm PT}/n_c$) becomes singular 
in the limit $n_{c}\rightarrow \infty$, 
for all the steps are subdivided and the widths of the steps shrink without limit as $n_{c}$ increases.
This is clear from Fig.~\ref{sb2}, in which 
$n(\alpha_0)\equiv N_{\rm PT}/n_c$ is plotted 
for $n_{c}=$100 and 500.
Note that $n(\alpha_0)$ is minimized at the golden angle. 
But, from the figure, 
it appears almost impossible to reach 
the golden angle continuously by
variational optimization.
In our static mechanism, 
the golden angle is 
singled out 
through the {screening process}.


%

\begin{figure}
\begin{center}
\includegraphics{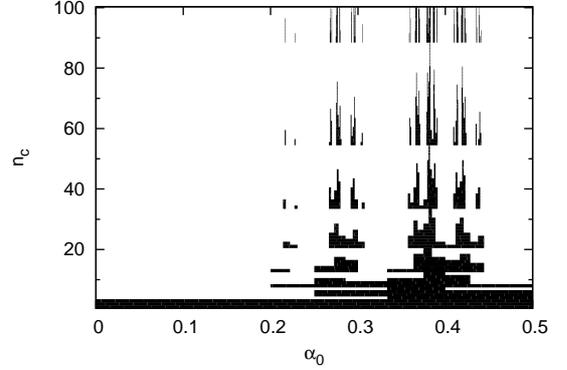}
\caption{\label{alpha-nummax}
The range of $\alpha_{0}$ within which 
$N_{\rm PT}$ takes the smallest value for given $n_{c}$
is filled with horizontal bars, thereby 
favorable values for $\alpha_{0}$ are pointed by `tapering needles'.
They are $\alpha_{0}\simeq 0.38, 0.28, 0.22$ (normal phyllotaxis), 
$\alpha_{0}\simeq 0.42, 0.44$,  and also  
$\alpha_{0}\simeq 0.30, 0.37$ (cf. Sec.~\ref{sec:majorseq}).
The golden angle $\alpha_0 \simeq 0.38$ ($2\pi\alpha_0 =137.5^\circ$) is singled out
for $n_c\simeq 20, 30, 50, 80$.
}
\end{center}
\end{figure}

\begin{figure}
\begin{center}
\includegraphics{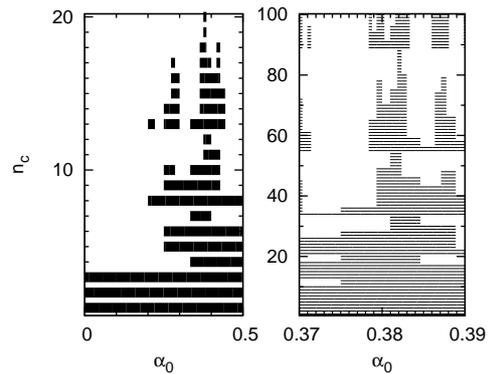}
\caption{\label{alpha-nummax2}
The range of $\alpha_{0}$ for the least $N_{\rm PT}$ is filled with
 horizontal lines 
(Fig.~\ref{alpha-nummax} enlarged 
vertically and horizontally).
}
\end{center}
\end{figure}

In Figs.~\ref{alpha-nummax} and \ref{alpha-nummax2}, 
the range of $\alpha_{0}$ within which the number $N_{\rm PT}$ is the smallest 
 for given $n_{c}$
is filled with the solid horizontal bar.
It is clearly shown that 
several values are specifically favored for $\alpha_{0}$.
Among others, we remark that the golden angle $2\pi \alpha_{0} \simeq 137.5^\circ$
($\alpha_{0} \simeq 0.382$)  is singled out 
for $n_{c}$ near but less than the Fibonacci numbers in (\ref{fibonacchi}).
%
%
The accuracy $\Delta\alpha$ for the golden angle converges rapidly,  
as mentioned above.
We have $\Delta \alpha=1/15$ for $n_{c}= 7$ in Fig.~\ref{sb4-7}, 
and $\Delta\alpha=1/1870$ around $n_{c}\simeq 85$ (cf. Table~\ref{tab:2111}).


To conclude, we may regard the hypothesis (H) as a law of phyllotaxis. 
In plain words,  the Fibonacci phyllotaxis with the main sequence  is favorably singled
out because of its special stability against inevitable structural changes  
 expected in the growing process.


\subsection{Mathematics}

The main results presented above 
are obtained directly 
from simple numerical analysis as shown in Fig.~\ref{pot5-7}.
Indeed it is straightforward to check them by hand when $n_{c}$ is small, 
but it soon gets complicated as $n_{c}$ becomes large.
Mathematical analysis helps us not only to derive useful formula 
but also to deepen understanding of the mathematical structure of the problem.
In particular, it is helpful for us to have analytical expressions of 
$(n_{0}, \Delta n)$ and $\Delta \alpha$ for 
PFs belonging to typical sequences found in nature.
The mathematical analysis is indispensable
if we do not content ourselves with several specific circumstances of lower phyllotaxis. 
Mathematical results are delegated to the appendixes.
Here we point out only that 
Fig.~\ref{tree} 
has the same structure as the {\it Stern-Brocot tree} 
of number theory 
(\cite{gkp94}), 
which contains each rational numbers exactly once.
%
%
Relations between numbers in the tree 
are concisely represented in terms of {\it mediants} (\ref{sec:mediant})
 and {\it continued fractions} (\ref{sec:confra}).
%
%

%
%

\subsection{Sequences}
\label{sec:majorseq}

As given in Table~\ref{alphanmax}, 
a general phyllotactic pattern derived from a random value of $\alpha_{0}$
does not fit in with observed regularity. 
Indeed, if PT would have to occur too frequently, 
the concepts of the PF and PT themselves could become indefinite.
As a matter of fact, fortunately, 
observed sequences
come around with astonishing regularity. 
Only several types of sequences exist in nature.
For the purpose of classifying the observed sequences, 
one often adopts a tacit theoretical procedure of 
inferring a mathematical limit $\alpha$ 
of a given sequence by extrapolation.
The limit divergence $\alpha$, generally an irrational number, 
 is then used to represent the sequence.
However, it must be kept in mind that
any phyllotactic sequence does terminate finitely in practice,
and inferring a limit from a finite sequence may be problematic.
Be that as it may, 
a sequence of principal convergents of 
a noble number 
(Eq.~(\ref{noblenumber}))
has been a central subject.
We can 
assess frequencies of occurrence of various sequences 
quantitatively by comparing $N_{\rm PT}$ for the sequences.
Below we use a shorthand bracket notation for a sequence
given in a paragraph below Eq.~(\ref{alpha2t111}) in \ref{sec:confra}.

The main sequence with the limit divergence $\alpha_0=0.382$ ($2\pi\alpha_0=137.5^\circ$) 
is given by 
\begin{equation}
[2]:  
1/2, 1/3, 2/5, 3/8, 5/13, 8/21, 13/34, 21/55, 
\cdots.
\label{[2]} 
\end{equation}
The first accessory sequence 
with the limit $\alpha_0=0.276$ ($2\pi\alpha_0=99.5^\circ$)
is 
\[
[3]:  
1/2, 1/3, 1/4, 2/7, 3/11, 5/18, 8/29, 13/47, 
\cdots. 
\]
The second accessory sequence for $\alpha_0=0.217$ ($2\pi\alpha_0=78.0^\circ$) 
is 
\[
[4]: 
1/2, 1/3, 1/4, 1/5, 2/9, 3/14, 5/23, 8/37,  
\cdots. 
\]
These are called normal phyllotaxis. 
As an example of anomalous phyllotaxis, 
the first lateral sequence 
for $\alpha_0=0.420$ ($2\pi\alpha_0=151.1^\circ$)
is 
\[
[2,2]: 
 1/2, 1/3, 2/5, 3/7, 5/12, 8/19, 13/31, 21/50, 
\cdots. 
\]
In addition, we can think of 
the sequence for $\alpha_0=0.296$ ($2\pi\alpha_0=106.4^\circ$), 
\[
[3,2]: 
 1/2, 1/3, 1/4, 2/7, 3/10, 5/17, 8/27, 13/44, 
\cdots.
\]
And, for $\alpha_0=0.367$ ($2\pi\alpha_0=132.2^\circ$), 
\[
[2,1,2]: 1/2, 1/3, 2/5, 3/8, 4/11, 7/19, 11/30, 18/49, \cdots. 
\]
There is controversy concerning the existence of the last two sequences 
(\cite{zm94,jean94}).
As the general fact of observation, 
any other sequence than the main sequence [2]
may be regarded as exceptional. 

\begin{figure}[t]
\begin{center}
\includegraphics{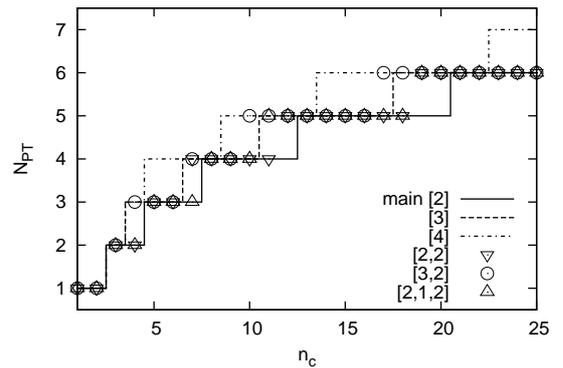}
\caption{\label{fig:numtr} 
$N_{\rm PT}$ is plotted against $n_{c}$ for major sequences. 
The sequences with the limit index 
$\alpha_0\simeq 0.38, 0.28, 0.22, 0.42, 0.30$ and 0.37
are represented as
[2], [3], [4], [2,2], [3,2] and [2,1,2], respectively. 
The main sequence [2] sets the lower limit of $N_{\rm PT}$. 
}
\end{center}
\end{figure}
For these sequences, 
$N_{\rm PT}$ is plotted against $n_{c}$ in Fig.~\ref{fig:numtr}.
The main sequence always sets the lower limit of $N_{\rm PT}$.
At $n_{c}=4$, [2,2] and [2,1,2] have $N_{\rm PT}=2$, 
and the others have $N_{\rm PT}=3$.
This is because 
the first three terms of [2,2] and [2,1,2] (up to 2/5) are the same as the main sequence [2].
From around $n_{c}=10$, 
the priority order of occurrence of the sequences
is inferred as  [2,2], [3], [3,2], and [4]. 
For such small $n_{c}$ as shown in Fig.~\ref{fig:numtr}, 
the order of [2,1,2] and [2,2]  cannot be decided uniquely. 
Note that 
[3,2] and [2,1,2] may be regarded as satellite sequences of [3] and [2],
respectively.
This is seen from Fig.~\ref{alpha-nummax}.
Therefore,  according to our result, 
extraordinary [3,2] and [2,1,2] are less unlikely
than [4] of `normal' phyllotaxis.
In Table~\ref{tab:nc}, for general sequences, 
we tabulate the values of $n_c$ at which $N_{\rm PT}$ 
is minimized.
Many sequences become favorable 
when $n_c$ is equal to a Fibonacci number in (\ref{fibonacchi}).
This is clear from Fig.~\ref{alpha-nummax}. 
We find that normal phyllotaxis of higher order 
[$n$] $(n=4,5,\cdots)$ appears not so specially preferable as
widely supposed.
In fact, Fig.~\ref{alpha-nummax} indicates that 
$N_{\rm PT}$ has 
no absolute minimum for $0<\alpha_0 <0.2$  and $n_c\ge 4$ 
($\alpha_0=0.178$ for [5] and $\alpha_0=0.151$ for [6]). 
A low-order sequence with [$l,m,n$] $(l,m,n\le 3)$ looks rather favorable. 

The order in the frequency of occurrence is generally consistent with
the available data, 
though the number of observations is still too limited to draw a definite
conclusion (\cite{fujita37,jean94}).


\begin{table*}[t]
\begin{center}
\caption{\label{tab:nc} 
Table for the growth index $n_c$ at which 
a given sequence makes a pattern with the minimum $N_{\rm PT}$. 
In the first row, the minimum value $N_{\rm PT}$ is given. 
The main sequence [2]: (1,2) covers all integers for $n_c$ ($N_{\rm PT}=2$ for $n_c=3,4$,
 etc.). 
The bracket notation for the sequence in the first column
is explained in a paragraph below Eq.~(\ref{alpha2t111}).
(A pair of numbers in the round bracket are parastichy pairs, i.e.,
 seed values for a Fibonacci recurrence relation. )
}
 \begin{tabular}{l|ccccccccc}
\hline
 $N_{\rm PT}$
&2&3&4&5&6&7&8&9 
\\
\hline 
  $[2]:  (1,2)$ & 3,4& 5-7& 8-12& 13-20& 21-33& 34-54& 55-88& 89-143  \\
  $[3]: (1,3)$  & 3& 5,6& 8-10& 13-17& 21-28& 34-46& 55-75& 89-122 \\
$[4] : (1,4)$  & 3& &8 &13 &21,22 &34-36 &55-59 & 89-96  \\ 
$[5] : (1,5)$  & 3& & & & & & &  \\ 
  $[2,2] : (2,5)$  &3,4 & 5,6& 8-11& 13-18& 21-30& 34-49& 55-80&89-130  \\ 
  $[2,3] : (2,7)$ & 3,4& 5,6& 8&13-15 &21-24 &34-40 &55-65 &89-106  \\ 
$[2,4] : (2,9)$   &3,4 &5,6 &8 & & & & &  \\ 
$[3,2] : (3,7)$   &3 &5,6 &8,9 &13-16 &21-26 &34-43 &55-70 &89-114  \\
$[3,3] : (3,10)$   &3 &5,6 &8,9 & &21,22 &34,35 &55-58 &89-94  \\ 
  $[3,4] : (3,13)$ & 3& 5,6& 8,9& & & & & \\ 
$[4,2] : (4,9)$  &3 & &8 & &21 &34 &55,56 &89-91  \\
$[4,3] : (4,13)$   &3 & &8 & & & & &  \\ 
$[2,1,2] : (3,8)$   &3,4 &5-7 &8-10 &13-18 &21-29 &34-48 &55-78 &89-127 
  \\ 
$[2,1,3] : (3,11)$   &3,4 &5-7 &8-10 &13 &21-24 &34-38 &55-63 &89-102  \\ 
$[2,1,4] : (3,14)$   & 3,4& 5-7& 8-10& 13& & & &  \\
$[2,2,2] : (5,12)$   &3,4 &5,6 &8-11 &13-16 &21-28 &34-45 &55-74 &89-120 
  \\ 
$[2,2,3] : (5,17)$   &3,4 &5,6 &8-11 &13-16 &21 &34-38 &55-60 &89-99  \\
$[2,2,4] : (5,22)$   &3,4 &5,6 &8-11 &13-16 &21 & & &  \\
$[2,3,2] : (7,16)$   &3,4 &5,6 &8 &13-15 &21,22 &34-38 &55-61 &89-100  \\
$[2,3,3] : (7,23)$   &3,4 &5,6 &8 &13-15 &21,22 & & &  \\
$[2,3,4] : (7,30)$   &3,4 &5,6 &8 &13-15 &21,22 & & &  \\
$[2,4,2] : (9, 20)$   &3,4 &5,6 &8 & & & & &  \\ 
$[3,1,2] : (4,11)$   &3 &5,6 &8-10 &13,14 &21-25 &34-40 &55-66 &89-107  \\
$[3,1,3] : (4,15)$   &3 &5,6 &8-10 &13,14 & & & &  \\
$[3,2,2] : (7 ,17)$   &3 &5,6 &8,9 &13-16 &21-23 &34-40 &55-64 &89-105  \\
$[3,2,3] : (7,24)$   &3 &5,6 &8,9 &13-16 &21-23 & & &  \\
$[3,3,2] : (10,23)$   &3 &5,6 &8,9 & &21,22 & &55 &  \\
$[3,3,3] : (10,33)$   &3 &5,6 &8,9 & &21,22 & & &  \\
$[4,1,2] : (5,14)$   &3 & &8 &13 & & & &  \\
$[4,1,3] : (5,19)$   &3 & &8 &13 & & & &  \\
$[4,2,2] : (9,22)$   &3 & &8 & &21 & & &  \\
\hline
 \end{tabular}
\end{center}
\end{table*}

\section{Discussion}
\label{sec:optimalwidth}

\begin{figure}
\begin{center}
\includegraphics{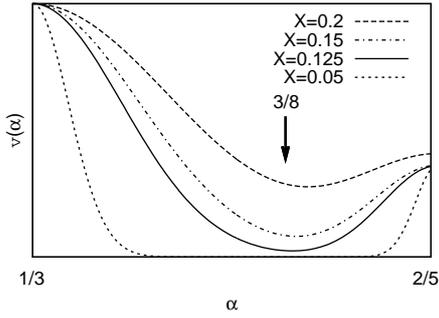}
\caption{\label{pot38}
For various $X$ around an optimal value $\bar{X}=0.125$, 
the effective potential $v(\alpha)$ 
is plotted by using Eq.~(\ref{V(x)}) and $v_n=0.6^n$ for $n\le n_{c}$ 
($5\le n_{c}<8$). 
For $X=0.05$ ($\ll \bar{X}$), 
$v(\alpha)$ has a flat bottom for a wide range of $\alpha$.
}
\end{center}
\end{figure}

We have made full use of 
an important result
of our model that 
the phyllotactic index $\alpha$ 
is given by  a fraction.
As a problem of macroscopic physics, 
it goes without saying that this is but a good approximation 
and mathematical rigor should not be expected in this respect.
In this section, 
we investigate the effect of the lateral width $X$ of 
the interaction $V(x)$ in Eq.~(\ref{V(x)}). 
There is an optimal value $\bar{X}$ for $X$ around which 
$\alpha$ is given by a fraction, as expected.
The optimal width $\bar{X}$ may be used 
to see if a minimum of  $v(\alpha)$ is really achieved practically. 
If the potential function $v(\alpha)$ is nearly constant and flat in a wide region around a minimum, 
it will take too long time to reach the true minimum 
because
the torsional driving force $- \frac{d v(\alpha)}{d\alpha}$ 
toward the minimum,
in the right-hand side of 
Eq.~(\ref{taudalphadt=-dedalpha}),  becomes practically zero. 
%
This happens if $X\ll \bar{X}$, 
as shown in Fig.~\ref{pot38} for $X=0.05$. 
It is known that 
the use of a fraction,
as originally made by Schimper and Braun, 
is not always adequate (\cite{fujita39}).

To put it concretely, 
let us examine PF 3/8, 
which occurs in between 1/3 and 2/5.
We consider 
\begin{equation}
1/3 < \alpha <2/5,
\label{13alpha25}
\end{equation}
and 
\begin{equation}
5\le n_{c} <8. 
\end{equation}
Then, for $v(\alpha)$ in Eq.~(\ref{valphaequiv}), 
all the other terms than $m=3$ and $m=5$ 
are effectively neglected, so that we get 
\begin{eqnarray}
 v(\alpha) &\simeq&  v_3 V(3 \alpha) +v_5 V(5 \alpha) 
\nonumber\\
&=& v_3 V(3 \alpha-1) +v_5 V(5 \alpha-2), 
\label{valphasimeqv3V3alphav5V5}
\end{eqnarray}
for $V(x)$ is periodic.
In effect, this is a minimal model for PF 3/8.
This may be interpreted as a mathematical expression of Hofmeister's rule, that new `leaf' arises in the largest gap between the previous ones. Note, however, that we are not concerned with mechanical dynamics of phyllotaxis at the apex. 

On the one hand, as a function of $\alpha$, 
$V (3 \alpha-1)$ has a peak 
with the width $X/3$ at $\alpha=1/3$, 
the lower boundary of (\ref{13alpha25}).
On the other hand, 
$V (5 \alpha-2)$ has a peak 
with the width $X/5$ at $\alpha=2/5$, 
the upper boundary of (\ref{13alpha25}).
An optimal width $\bar{X}$ is estimated 
by equating the total width $ \bar{X}/3 +  \bar{X}/5$
with the allowed range for $\alpha$, that is, $\Delta \alpha=2/5-1/3$.
We obtain the result $\bar{X}=1/8$ for 3/8. 
According to Eq.~(\ref{Deltathetasimeq4piX}), 
the optimal angular width of interaction  is 
$\Delta \theta\simeq 90^\circ$ for 3/8.
In the ideal case $X=\bar{X}$, 
$v(\alpha)$ has a minimum at
$\alpha=1/3 + \bar{X}/3= 2/5- \bar{X}/5=3/8$, 
as expected.
The derivation outlined here 
indicates that the result 
will not depend on $V(\alpha)$ specifically. 
%

We present Fig.~\ref{pot38} to show the effect of 
$X$ on $v(\alpha)$.
To draw the figure, 
we use a full form of $v(\alpha)$ and
did not use the approximation in
Eq.~(\ref{valphasimeqv3V3alphav5V5}). 
In any case, the difference due to the approximation is negligible. 
As mentioned, 
for $X\simeq \bar{X}$, 
we obtain $\alpha\simeq 3/8$ properly with good accuracy.
As $X$ is decreased from $\bar{X}$, $v(\alpha)$ around the minimum gets flattened.
As a matter of practical fact, 
the minimum would not be reached in the extreme case of $X\ll \bar{X}$. 
When $X<\bar{X}$, 
there appears a flat region in
\begin{equation}
\frac{1}{3} + \frac{X}{3} < \alpha< \frac{2}{5} - \frac{X}{5}, 
\end{equation}
the width of which is 
$\Delta \alpha_{\rm flat}=(1-8X)/15
=(1-\frac{X}{\bar{X}})/15$. 
This is smaller than the full width $\Delta \alpha=1/15$ of (\ref{13alpha25}),  as it should be.
In practice,  
insofar as $\alpha$ stays within the flat region, 
we may rather observe
the primary angle $\alpha=\alpha_{0}$ as it is, 
since the secondary torsion is not effective any longer.

In general, 
the optimal width $\bar{X}$ for 
a fraction $p/q$ is
given by $\bar{X}=1/q$, 
and 
the width of the flat region is 
\begin{equation}
 \Delta \alpha_{\rm flat}= 
\left(1-\frac{X}{\bar{X}}\right)\Delta \alpha,
\label{Deltaalphaflat}
\end{equation}
for $X<\bar{X}$  (\ref{sec:dynamicalmodel}). 
As we follow any sequence up in the tree of Fig.~\ref{tree},
the denominator $q_n$ of PF $\alpha=p_n/q_n$ stays constant or increases, 
so that $\bar{X}$ is constant or decreases. 
In effect,  it decreases roughly as $\bar{X}\sim 1/n_{c}$.
By contrast, 
the model parameter $X$ 
should be fit with a real plant. 
Consequently, 
the condition $X<\bar{X}$ may hold in early few terms of a sequence
(namely, 1/2, 1/3, etc., when $n_{c}$ is small). 
Then we could no more expect to observe these low order PFs. 
We rather observe an inherent value $\alpha=\alpha_{0}$
so far as it falls within a flat region with the width $\Delta \alpha_{\rm flat}$.
This is consistent with observations. 

%

%
%
%
%

%
%
%

%
%
%
%
%
%
%

\section{Conclusions}

With the aid of {biological} hypotheses,
we showed that prevalent sequences of  phyllotactic fractions
are satisfactorily explained by a {physical} model of plant growth.
The model has interesting {mathematical} properties. 
Among others,  
we bring to light a Stern-Brocot type 
number-theoretical structure that has been unnoticed thus far in this field. 
To extract the mathematical essence of the phenomenon, 
we have to base our theory on the abstract model 
by discarding real biological implementation as non-essential details.
According to the proposed static mechanism, 
the phyllotactic pattern with the main Fibonacci sequence is naturally selected
because it entails the least number of phyllotactic structural transitions
while growing to a mature plant. 


%
%
%
%
%
%
%
%
%
%
%

%
\section*{Acknowledgement}
I wish to express my appreciation to the reviewer for 
the invaluable 
comments which helped me improve the paper significantly.
%

%

\bibliographystyle{elsarticle-harv}
\bibliography{okabe}

\appendix

\section{Mediant}
\label{sec:mediant}

The {\it mediant} of two fractions $m/n$ and $p/q$ is given by
$(m+p)/(n+q)$, 
where $m$ and $n$ ($p$ and $q$) are relatively prime integers. 
Let us call $m/n$ and $p/q$ as parent fractions of 
a child fraction $(m+p)/(n+q)$. 

The Stern-Brocot tree
of fractions  between 0 and 1/2 is obtained by the following operation (\cite{gkp94}). 
Start from the initial fractions 0/1(=0) and 1/2. 
Repeat 
inserting the mediant of two adjacent fractions, 
and arranging them in numerical order.
The first mediant is 1/3 between 0/1 and 1/2, 
and they are arranged as (0/1), 1/3, (1/2). 
In the second order, we obtain 1/4 between 0/1 and 1/3, 
and 2/5 between 1/3 and 1/2.  
They are arranged as (0/1), 1/4, (1/3), 2/5, (1/2). 
In the third order, 
we obtain (0/1), 1/5, (1/4), 2/7, (1/3), 3/8, (2/5), 3/7, (1/2).
Here we put the fractions in the previous orders in parentheses.
The Stern-Brocot tree has been anticipated by Schimper
(\cite{schimper1835beschreibung,hn10}).

Our tree in Fig.~\ref{tree} is related to but not the same as the Stern-Brocot tree. 
As an important difference,  we have to order fractions by the growth
index $n_{c}$.
In other words, we need to know $(n_{0}, \Delta n)$ of the fractions.
For instance, let us consider $2/5$. 
For $n_{c}=3$ in Fig.~\ref{tree}, 
$(m+p)/(n+q)=2/5$ is 
the mediant of $p/q=1/3$ and $m/n=1/2$ ($m=1, n=2, p=1,q=3$). 
By ordering the fractions according to the denominators, $q$ and $n$, 
let us call $p/q=1/3$ and $m/n=1/2$  as the younger and the older parent of 2/5.
On the one hand, the mediant 2/5 is born 
when the younger parent $p/q=1/3$ dies at $n_{c}=q=3$.
On the other hand, 
2/5 dies at $n_{c}=n+q=5$, the denominator of 2/5.
Therefore,  we obtain $3 \le  n_{c} < 5$ and $\Delta n=1$ for 2/5.
In general, 
the mediant $(m+p)/(n+q)$ of $m/n$ and $p/q$ ($n<q$) has $q \le  n_{c} < n+q$.
For $(m+p)/(n+q)$ ($n<q$), we obtain 
\begin{equation}
(n_{0}, \Delta n)=(q, n-1). 
\label{apn0deltanqn-1}
\end{equation}

Next we consider $\Delta \alpha$. 
In order to realize $\alpha=(m+p)/(n+q)$, 
$\alpha_{0}$ has to lie between the parents $m/n$ and $p/q$, 
and we get $\Delta \alpha=|\frac{m}{n} - \frac{p}{q}|$. 
The parent fractions belonging to the tree are shown to satisfy 
$|mq-np|=1$ (\cite{gkp94}). 
Hence 
we conclude
\begin{equation}
\Delta \alpha=\frac{1}{nq}
\label{delalphaini}
\end{equation}
for $(m+p)/(n+q)$.
These formulas may be used for Fig.~\ref{deln-alini}. 

\begin{figure}
\begin{center}
\includegraphics[width=.5\textwidth]{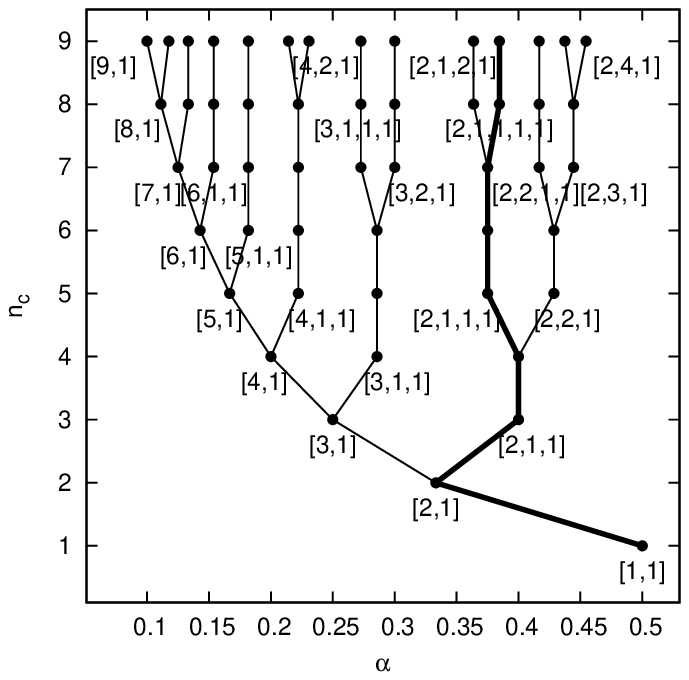}
\caption{\label{branch}
The branching structure in Fig.~\ref{tree} is schematically shown 
(a) for $p/q < m/n$ and (b) for $p/q > m/n$, 
where 
$m$ and $n$, $p$ and $q$ are relatively prime numbers satisfying $n<q$.
In both cases,  two daughter fractions 
$(m+2p)/(n+2q)$ and $(2m+p)/(2n+q)$ 
at the top of the figure are derived from a mother fraction $(m+p)/(n+q)$ in the middle.
The mother $(m+p)/(n+q)$ is derived as a child of the parents $p/q$ and $m/n$. 
We obtain $\Delta n=q-1$ for $(m+2p)/(n+2q)$, and 
$\Delta n= n-1$ for $(2m+p)/(2n+q)$ and $(m+p)/(n+q)$. 
The path to increase $\Delta n$ traces 
the bold zigzag line. 
}
\end{center}
\end{figure}
For the hypothesis (H) in Sec.~\ref{subsec:main}, 
we have to compare $\Delta n$ of 
 two daughter fractions derived from a mother fraction in
Fig.~\ref{tree}. 
(We tell a mother-daughter relation from a parent-child relation.)
Consider $(m+p)/(n+q)$ as a mother fraction,  derived 
as a child of parents $m/n$ and $p/q$ ($n<q$) (Fig.~\ref{branch}).
One daughter fraction $(2m+p)/(2n+q)$ 
occurs between $(m+p)/(n+q)$ and $m/n$. 
The other daughter fraction 
$(m+2p)/(n+2q)$ 
occurs between $(m+p)/(n+q)$ and $p/q$.
On account of Eq.~(\ref{apn0deltanqn-1}), 
the former $(2m+p)/(2n+q)$ has $(n_{0}, \Delta n)=(n+q , n-1)$, 
whereas the latter $(m+2p)/(n+2q)$ has $(n_{0},\Delta n)=(n+q, q-1)$. 
By assumption $n<q$, 
the daughter fraction 
$(m+2p)/(n+2q)$ has the larger $\Delta n=q-1$ than 
$(2m+p)/(2n+q)$ with $\Delta n=n-1$.
{\it For two daughter fractions derived from a mother fraction, 
the one with a larger denominator always increases $\Delta n$,
whereas the other does not change $\Delta n$ from the mother fraction.}

This rule is read from Fig.~\ref{deln-alini}. 
At every branching point (node),  one branch grows to the right to
increase $\Delta n$. 
The other goes down along the ordinate in the main figure.
According to (H), 
the sequence with ever increasing $\Delta n$
comprises the most favorable branch of our evolutionary tree.
A favored sequence in the tree diagram of Fig.~\ref{tree} traces 
 a zigzag path as depicted with the bold line
in Fig~\ref{branch}.
This result may be compared with a dynamical counterpart 
 (\ref{sec:confra}, \ref{sec:dynamicalmodel}). 
%
%
%



\begin{figure}[t]
\begin{center}
\includegraphics{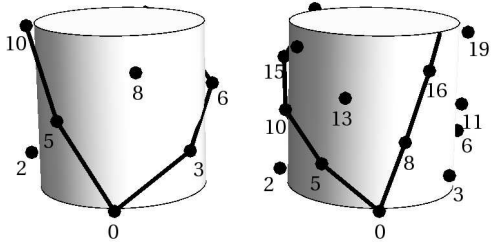}
\caption{\label{para813}
Patterns with the same divergence angle 
$\alpha=0.38$ appear
to have different parastichy pairs $(n,q)=$
(3,5) (left) and (5,8) (right), depending on a vertical length scale.
The 5-parastichy winds up clockwise.
Thus this `transition' of a parastichy pair is a superficial phenomenon.
The apparent change from (3,5) to (5,8) is 
called rising phyllotaxis.
Rising phyllotaxis restricts a range allowed for $\alpha$ (Fig.~\ref{treePS}). }
\end{center}
\end{figure}

\begin{figure}[t]
\begin{center}
\includegraphics[width=.5\textwidth]{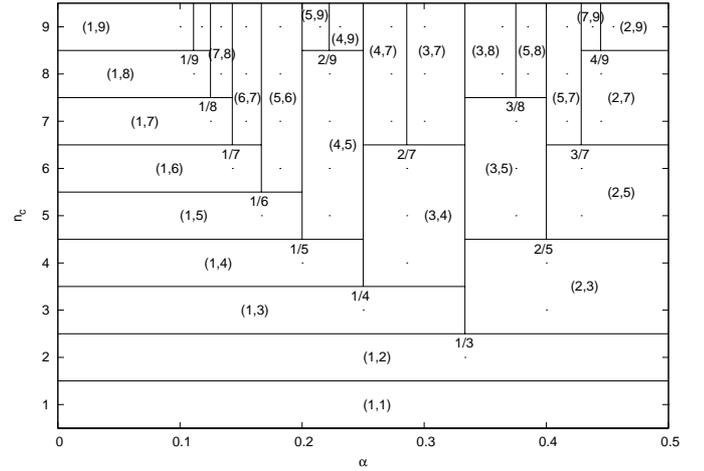}
\caption{\label{treePS}
Diagram of a phyllotactic system with one genetic spiral, showing the visible opposed parastichy pair $(n,q)$.
For $3/8<\alpha< 2/5$, we may obtain 
(1,2), (2,3), (3,5) or  (5,8), depending on a vertical length scale 
(Fig.~\ref{para813}). 
Conversely, if we observe rising phyllotaxis through parastichy pairs
 (1,2), (2,3), (3,5) and (5,8),  then 
the divergence angle $\alpha$ must meet the constraint 
$3/8<\alpha< 2/5$ $(\Delta \alpha=0.025)$. 
For a system with $J$ genetic spirals, $\alpha$ and $(n,q)$ are replaced by $J\alpha$ and $J(n,q)$, respectively.
}
\end{center}
\end{figure}
In practice, 
when the divergence angle $\alpha$ deviates from an exact fraction, 
a pair of integers  $(n,q)$, called a {\it parastichy pair}, 
is used to represent a phyllotactic pattern consisting of 
most conspicuously 
visible families of $n$ and $q$ secondary spirals (parastichies) 
crossing with each other. 
A $q$-parastichy is a secondary spiral running through the points $(p, \theta_p)$, 
$(p+q, \theta_{p+q})$,  $({p+2q}, \theta_{p+2q})$,  
$({p+3q}, \theta_{p+3q})$, etc., where $p=0,1,2,\cdots, q-1$ (Fig.~\ref{para813}).
To derive a parastichy pair for given $\alpha$ 
is a purely geometrical problem (\cite{adler74,jean94}). 
In our tree system, 
we find a simple result that
{\it we obtain a visible parastichy pair $(n,q)$ 
when $\alpha$ lies between two neighboring (parent) fractions $m/n$ and $p/q$ 
in our tree.}
Accordingly, PF $\alpha=(m+p)/(n+q)$ may be replaced by 
the parastichy pair $(n,q)$. 
In place of Fig.~\ref{treePF}, 
we obtain Fig.~\ref{treePS}, which may be useful
to analyze real systems.
%



\section{Continued Fraction}
\label{sec:confra}

A real number $\alpha$ $(0<\alpha < 1)$ is represented as
a {\it continued fraction}, 
\begin{equation}
 \alpha =\frac{1}{a_1+\dfrac{1}{a_2+ \dfrac{1}{a_3+\cdots}}} \equiv [a_1, a_2, a_3,\cdots],
\label{alpha=frac1/a1+1a21a3}
\end{equation}
where $a_i$  $(i=1,2,\cdots)$ is an positive integer. 
Every rational number has two continued fraction expansions.
In one the final term is 1, that is,  a rational number $\alpha$ is 
represented finitely as $\alpha=[a_1, a_2, \cdots, a_n,1]$. 
For an irrational number $\alpha$, 
there is a successive rational approximation, 
${p_n}/{q_n} = [a_1, \cdots, a_n]$ ($n=1,2,\cdots$), 
in terms of relatively prime positive integers $p_n$ and $q_n$ satisfying 
the recursion relations, 
\begin{eqnarray}
 p_{n+2} &= & a_{n+2}p_{n+1} +p_n,
\\
 q_{n+2} &= & a_{n+2}q_{n+1} +q_n,
\label{qn+2=an+2qn+1+qn}
\end{eqnarray}
and 
\begin{eqnarray}
 p_0=0,\ p_1=1,\ q_0=1,\ q_1=a_1.
\end{eqnarray}
The fraction $p_n/q_n$ $(n=1,2,\cdots)$ is 
called a {\it principal convergent} of $\alpha$. 
The difference between successive principal convergents is given by 
\begin{equation}
\frac{p_{n+1}}{q_{n+1}}  -\frac{p_n}{q_n}  
=
\frac{(-1)^{n}}{q_nq_{n+1}}, 
\label{pn+1qn+1-pnqn=}
\end{equation}
and $\alpha$ lies 
between 
even and odd order convergents,
\begin{equation}
 \frac{p_{2k}}{q_{2k}}<\alpha<  \frac{p_{2k+1}}{q_{2k+1}}.
\label{p2kq2k<alphap2k+2}
\end{equation}
Thus
the principal convergent $p_n/q_n$ approaches to the limit $\alpha$
in a zigzag manner. 
\begin{figure}
\begin{center}
\includegraphics{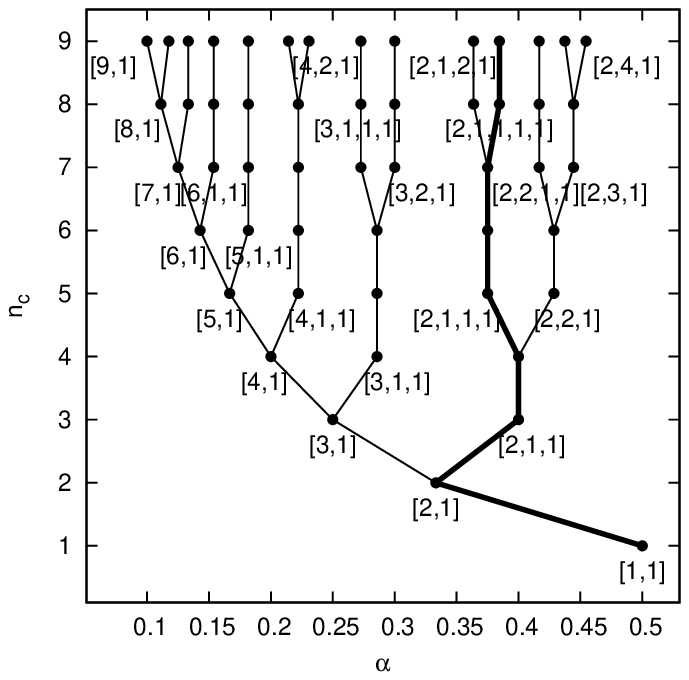}
\caption{\label{treeCF}
Fractions in Fig.~\ref{tree} are represented with 
the bracket notation for
 continued fractions defined in Eq.~(\ref{alpha=frac1/a1+1a21a3}).  
}
\end{center}
\end{figure}


In terms of the bracket notation 
defined in Eq.~(\ref{alpha=frac1/a1+1a21a3}),  we obtain
Fig.~\ref{treeCF} in place of Fig.~\ref{tree}. 
From the figure,  we immediately notice the bifurcation rule holding at every node, 
\begin{equation}
 [a_1,a_2, \cdots,a_n, 1]
\longrightarrow
\begin{array}{c}
 [a_1,a_2, \cdots,a_n, 1,1] \\

 [a_1,a_2, \cdots,a_n+1, 1]. \\
\end{array}
\label{aprulea1a2cdots}
\end{equation}
It is easily checked that 
the upper branch in (\ref{aprulea1a2cdots}) increases $\Delta n$, 
and the lower one conserves $\Delta n$ (\ref{sec:mediant}). 
Therefore, 
according to (H) in Sec.~\ref{subsec:main}, 
it is particularly important for us to study 
the following sequence of fractions ($a_i\ne 1$ for $i=1,2,\cdots, n$): 
\begin{eqnarray}
&&[a_1,a_2, \cdots, a_n, 1], \nonumber\\ 
&&[a_1,a_2, \cdots, a_n, 1,1], \nonumber\\ 
&&[a_1,a_2, \cdots, a_n, 1,1,1], \nonumber\\ 
&&[a_1,a_2, \cdots, a_n, 1,1,1,1],  \cdots.
\end{eqnarray}
These are comprised in the {principal convergents} of a {\it noble number}, 
\begin{equation}
\alpha_{\rm noble}= [a_1,a_2, \cdots,a_n, 1,1,1,1, \cdots].  
\label{noblenumber}
\end{equation}
Owing to the succession of $a_i=1$ $(i>n)$, 
{\it 
for a sequence of fractions $p_n/q_n$ on a favored branch according to (H),  
the numerator $p_n$ and denominator $q_n$ 
obey the Fibonacci recursion relations $p_n=p_{n-1}+p_{n-2}$ and $q_n=q_{n-1}+q_{n-2}$. }

A sequence of  principal convergents of a noble number 
has been given a special status in theoretical studies of phyllotaxis (\cite{coxeter72,mk83,ropl84,rk89a}). 
The most important is the golden angle (per $2\pi$)
for $a_1=2$ and $a_n=1$ $(n=1,2,3,\cdots)$, 
\begin{equation}
 \alpha=[2, 1,1,1,\cdots] = 1/(2+\tau^{-1}) = \tau^{-2} =  (3-\sqrt{5})/2, 
\label{goldenalpha}
\end{equation}
where the golden ratio $\tau$ is given by 
\begin{equation}
 \tau=1+ [1,1,1,\cdots] = (\sqrt{5}+1)/2, 
\end{equation}
or
\begin{equation}
 \tau^{-1}= [1,1,1,\cdots] = (\sqrt{5}-1)/2.
\end{equation}

In the literature, a sequence is referred to in several ways.
In particular, a sequence with a limit index 
\begin{equation}
 \alpha=[t, 1,1,1,\cdots] = 1/(t+\tau^{-1})
\label{normalphylalpha}
\end{equation}
($t=2,3,\cdots$) is called {\it normal phyllotaxis}.
The main sequence for Eq.~(\ref{goldenalpha})
corresponds to the special case $t=2$. 
The cases $t=3$ and $t=4$ 
are called the first and the second {\it accessory sequence}, respectively.
On the other side, a sequence with a limit angle
\begin{equation}
 \alpha=[2, t, 1,1,1,\cdots] = 1/(2+ (t+\tau^{-1})^{-1}), 
\label{alpha2t111}
\end{equation}
($t=2,3,\cdots$) 
is sometimes called the {\it lateral sequence}.

For convenience sake, let us introduce another concise notation.
To denote a whole sequence with a limit of 
\begin{equation}
 \alpha=[a_1,a_2, \cdots,a_n, 1,1,\cdots]
=1/(a_1+(a_2+(\cdots
+(a_n+\tau^{-1})^{-1}
)^{-1}
)^{-1}), 
\end{equation}
we simply reuse a symbol for a continued fraction 
$[a_1,a_2, \cdots,a_n]$ ($a_n\ne 1$).  
Hence  the main sequence is 
\begin{equation}
[2]:  1/2, 1/3, 2/5, 3/8, 5/13, 8/21, 13/34, 21/55, 
\cdots.
\end{equation}
To obtain the sequence $[3,2]$ with a limit $\alpha=1/(3+
(2+\tau^{-1})^{-1})$, 
collect the fractions from 1/2 to $[3,2]=2/7$
up along the tree in Fig.~\ref{tree},  
namely, 1/2, 1/3, 1/4, 2/7. 
Then, 
from the daughter fractions 3/10 and 3/11 of 2/7, 
choose the {unfavorable} one with a smaller denominator, 
namely,  3/10.
After them follow all the fractions on the favorable branch ramifying from
3/10, namely, 5/17, 8/27, 13/44, 21/71, 34/115, 55/186, 89/301, $\cdots$. 
As noted below Eq.~(\ref{noblenumber}), 
these obey the Fibonacci recursion relation 
(e.g., 13/44 = (5+8)/(17+27)). 
To summarize, we obtain 
\[
 [3,2]: 1/2, 1/3, 1/4, 2/7, 3/10,  5/17, 8/27, 13/44, 21/71, 34/115,
\cdots. 
\]
Major sequences  are given in Sec.~\ref{sec:majorseq}.

A sequence is commonly represented by parastichy numbers instead of fractions. 
Translation is made without difficulty 
by noticing the denominators (\ref{sec:mediant}).
\begin{equation}
[2]:  (1,1), (1,2), (2,3) ,(3, 5), (5 ,8), (8,13), (13, 21),   \cdots.
\end{equation}
\begin{equation}
[3,2]:  (1, 1), (1,2), (1,3), (3,4), (3,7), (7,10), (10,17), (17,27), 
\cdots.
\end{equation}
We may combine repeated numbers on favored branches. 
\begin{equation}
[2]:  1, 2,3, 5 ,8, 13, 21,   \cdots.
\end{equation}
\begin{equation}
[3,2]:  (1,2), (1,3,4), (3,7,10,17,27, 
\cdots.
\end{equation}
All but the last branch may be omitted. 
\begin{equation}
[3,2]:  3,7,10,17,27, 
\cdots.
\end{equation}
Most concisely, only the first two numbers 
 may be given as seed values of recurrence. 
\begin{equation}
[2]:  (1, 2). 
\end{equation}
\begin{equation}
[3,2]:  (3,7). 
\end{equation}
In this notation, 
typical sequences are generally given as follows (cf. Table~\ref{tab:nc}). 
\begin{equation}
  [a]:  (1,a).  \quad (a =2,3,\cdots.) 
\end{equation}
\begin{equation}
 [p,a]:  (p, ap+1).  \quad (a,p =2,3,\cdots.)
\end{equation}
\begin{equation}
 [p-1,1,a-1]:   (p, ap-1).  \quad (a,p =3,4,\cdots.)
\end{equation}
\begin{equation}
 [p-1,2,a-1]:  (2p-1, a (2p-1)-p).  \quad (a,p =3,4,\cdots.) 
\end{equation}
According to \cite{fujita37}, the first sequence system $[a]$ is common, 
the second sequence system $[p, a]$ is rarely observed, and 
the third sequence system $[p-1, 1, a-1]$ is extremely rare. 
This is consistent with our results in Table~\ref{tab:nc}.

%


\begin{table*}[t]
\begin{center}
\caption{\label{tab:2111}
The main sequence with the limit divergence angle 
$\alpha=\tau^{-2}$.
}
\begin{tabular}{ccccccccccc}
\hline
$n$ &1&2&3&4&5&6&7&8&9&10\\
\hline
$\dfrac{p_n}{q_n\strut }$   & $\dfrac{1}{2}$ & $\dfrac{1}{3}$ &
		 $\dfrac{2}{5}$ &$\dfrac{3}{8}$ & $\dfrac{5}{13}$
			 &$\dfrac{8}{21}$ &$\dfrac{13}{34}$
			     &$\dfrac{21}{55}$ &$\dfrac{34\strut}{89}$ &$\dfrac{55\strut}{144}$ 
\\
 \hline
 $n_{0}$ & 1 & 2 & 3 &5 &8 &13 & 21 & 34&55 &89\\ 
 $\Delta n$
 &0 & 0 &1 &2 &4 &7 &12 & 20&33&54 \\ 
\hline
 $\Delta \alpha$ & 1 & $\dfrac{1}{2}$  & $\dfrac{1}{6}$  &
		 $\dfrac{1}{15}$ & $\dfrac{1}{40}$ & $\dfrac{1}{104}$
			 &$\dfrac{1}{273}$ &$\dfrac{1}{714}$
				 &$\dfrac{1\strut}{1870\strut }$ &$\dfrac{1\strut}{4895\strut }$\\ 
\hline
 $\bar{X}$ & $\dfrac{1}{2}$ & $\dfrac{1}{3}$ &$\dfrac{1}{5}$ &
		 $\dfrac{1}{8}$ & $\dfrac{1}{13}$ & $\dfrac{1}{21}$ &
			     $\dfrac{1}{34}$& $\dfrac{1}{55}$&
				     $\dfrac{1\strut}{89\strut}$& $\dfrac{1\strut}{144\strut}$
\\ 
\hline
\end{tabular}
\end{center}
\end{table*}
Now our task is to get $\Delta \alpha$ and 
$(n_{0}, \Delta n)$ for a sequence of principal convergents of a
noble number.
From Eq.~(\ref{pn+1qn+1-pnqn=}), we get
\begin{equation}
\frac{p_{n}}{q_{n}}  -\frac{p_{n-1}}{q_{n-1}}  
=
\frac{(-1)^{n-1}}{q_{n-1}q_{n}}, 
\end{equation}
and 
\begin{equation}
\frac{p_{n}}{q_{n}}  -\frac{p_{n-2}}{q_{n-2}}  
=
\frac{(-1)^{n} a_n}{q_nq_{n-2}}. 
\end{equation}
These equations signify that 
$p_n/q_n$ lies between $p_{n-1}/q_{n-1}$ and  $p_{n-2}/q_{n-2}$.
We find that  $p_n/q_n$ has 
\begin{equation}
\Delta \alpha= 
\left|
\frac{p_{n-1}}{q_{n-1}}-
\frac{p_{n-2}}{q_{n-2}}
\right|=
\frac{1}{q_{n-1}q_{n-2}}.
\label{deltaalphaini1qi-1qi-2}
\end{equation}
According to  our model, a child fraction $p_n/q_n$ 
is born at $n_{c}=q_{n-1}$ from a parent fraction $p_{n-1}/q_{n-1}$, 
and dies at $n_{c}=q_{n}$.
Hence, we obtain 
$q_{n-1}\le  n_{c} < q_n$ for $p_n/q_n$. 
Using Eq.~(\ref{qn+2=an+2qn+1+qn}), 
for $p_n/q_n$ of a noble number with $a_{n}=1$, 
we conclude 
\begin{equation}
(n_{0}, \Delta n) =(q_{n-1}, q_{n-2}-1). 
\label{deltanmax=qi-2-1}
\end{equation}

\begin{table*}[t]
\begin{center}
\caption{\label{tab:fibo} Fibonacci sequence.\strut}
 \begin{tabular}{c|ccccccccccccc}
\hline
  $n$& 0 & 1&2 &3 &4 &5 &6 &7 &8 &9 &10&11&12 \\
\hline
  $F_n$& 0 & 1& 1& 2&3 &5 &8 &13 &21 &34 &55 &89&144 \\
\hline
 \end{tabular}
\end{center}
\end{table*}
To put it more concretely,  hereafter we restrict ourselves 
to the most important case of the golden angle in Eq.~(\ref{goldenalpha}). 
The principal convergent $p_n/q_n$, 
$\Delta \alpha$ and $(n_{0}, \Delta n)$ 
are presented in Table~\ref{tab:2111}.
In this simplest case, we have 
$p_n=F_{n}$ and $q_n=F_{n+2}$, 
where $F_n$ is the Fibonacci number defined by the recurrence 
\begin{eqnarray}
F_0&=&0, \nonumber\\
F_1&=&1, \nonumber\\
F_{n}&=& F_{n-1} +F_{n-2}.\qquad (n>1 ) \nonumber
\end{eqnarray}
(Table~\ref{tab:fibo}.)
The number of phyllotactic transition 
$N_{\rm PT}$ simply 
counts the number of $p_n/q_n$. 
Consequently, 
we obtain $N_{\rm PT}= 2n+1$
for 
\begin{equation}
\frac{p_{2n}}{q_{2n}}
<\alpha<
\frac{p_{2n+1}}{q_{2n+1}}
\label{p2nq2n<alphap}
\end{equation}
%
and 
\begin{equation}
q_{2n+1}\le  n_{c} < q_{2n+2}. 
\end{equation}
In terms of $F_n$, 
we obtain
$N_{\rm PT}= 2n+1$
for 
$
F_{2n+3}\le  n_{c} < F_{2n+4}, 
$
and 
$
\Delta \alpha= 1/({F_{2n+2}F_{2n+3}}).
$
As a result, 
irrespective of whether $n$ is even or odd, we obtain
\begin{equation}
 N_{\rm PT}= n
\end{equation}
for 
\begin{equation}
F_{n+2}\le  n_{c} < F_{n+3}, 
\label{Fn+3nc}
\end{equation}
and
\begin{equation}
\Delta \alpha= \frac{1}{F_{n+1}F_{n+2}}.
\label{deltaalpha=1fn+1}
\end{equation}

Using a formula (\cite{gkp94})
\begin{equation}
 F_n =\frac{1}{\sqrt{5}}\left(\tau^n- (-\tau)^{-n}\right)
\simeq \frac{\tau^n}{\sqrt{5}}, 
\end{equation}
we may regard 
$n_c\simeq {\tau^{N_{\rm PT}+2.5}}/{\sqrt{5}}$ by (\ref{Fn+3nc}). 
Then, 
we get
\begin{equation}
 N_{\rm PT}=\log (\sqrt{5} n_c) /\log \tau- 2.5,
\label{NPTlogsqrt5nc}
\end{equation}
and 
\begin{equation}
 \Delta \alpha= (\tau/n_c)^{2}. 
\label{Deltaalpha=taunc^2}
\end{equation}
Thus $N_{\rm PT}/n_c$ and $\Delta \alpha$ vanish in the limit
$n_c\rightarrow \infty$. 
The logarithmic dependence in Eq.~(\ref{NPTlogsqrt5nc}) 
for the irrational number $\alpha=\tau^{-2}$
is contrasted
with a linear dependence $N_{\rm PT}\simeq n_c/n$ 
for a rational number $\alpha=1/n$ (Fig.~\ref{sb2}).
These results are used to guess 
a growth index $n_c$ inversely.
To achieve accuracy of $\Delta \alpha=0.02$ (i.e., $\alpha=0.38\pm
0.01$), 
we need 
$n_c\simeq 11$ according to Eq.~(\ref{Deltaalpha=taunc^2}). 
To obtain a parastichy pair $(F_{10}, F_{11})=(55,89)$ of a sunflower
head, 
we have to attain 
$\Delta \alpha=\frac{1}{F_{10}F_{11}}=1/4895$,
for which we need
$n_c\simeq 113$. 
Indeed this lies between $F_{11}\le n_c <F_{12}$.
\section{Deviation from Fraction}
\label{sec:dynamicalmodel}

To generalize 
the discussion in Sec.~\ref{sec:optimalwidth}, 
let us consider a minimum 
of $v(\alpha)$ around a mediant 
\begin{equation}
\bar{\alpha}=\frac{m+p}{n+q}
\label{alpha0}
\end{equation}
of reduced fractions $m/n$ and $p/q$ (\ref{sec:mediant}). 
As in Eq.~(\ref{valphasimeqv3V3alphav5V5}),  
the minimal model for this purpose
is given by 
\begin{equation}
v(\alpha)=v_n  V (n \alpha-m)+v_q  V (q \alpha-p).  
\end{equation}
Differentiating this with respect to $\alpha$, 
and substituting Eq.~(\ref{alpha0}), 
we obtain 
\begin{equation}
v'\left(\bar{\alpha}
\right)= n v_n  V' \left(
\frac{np-mq}{n+q}
\right)+qv_q  V' \left(
\frac{qm-np}{n+q} 
\right). 
\end{equation}
On physical grounds, it is natural to assume that $V(x)$ is an even function,  $V(x)=V(-x)$.
Then, because of  $V'(x)=-V'(-x)$, 
\begin{equation}
v'\left(
\bar{\alpha}
\right)= 
(qv_q-n v_n
 )  V' \left(
\frac{mq-np}{n+q}
\right). 
\end{equation}
By definition, 
$V(\alpha)= V'(\alpha) =0$ for $\alpha> X$. 
Hence, 
we get $v'(\bar{\alpha})=0$ for $X< \bar{X}$, 
where $\bar{X}={|mq-np|}/{(n+q)}$.
Without  loss of generality, we may assume
\begin{equation}
\frac{m}{n} < \frac{p}{q}. 
\label{mn<pq}
\end{equation}
Using $mq-np=1$ (\cite{gkp94}), 
%
%
we obtain 
\begin{equation}
\bar{X}=\frac{1}{n+q} 
\label{barX1/n+q}
\end{equation}
for $({m+p})/({n+q})$. 
{\it The optimal width $\bar{X}$ for a fraction $p/q$ is
given by $\bar{X}=1/q$.}
For $X<\bar{X}$, 
the potential $v(\alpha)$ is nearly constant for
\begin{equation}
\frac{m}{n} + \frac{X}{n} < \alpha< \frac{p}{q} - \frac{X}{q}, 
\label{mnXnalphapqXq}
\end{equation}
the width of which is 
\begin{equation}
 \Delta \alpha_{\rm flat}= 
\left(1-\frac{X}{\bar{X}}\right)\Delta \alpha, 
\label{deltaflat}
\end{equation}
where we used 
Eqs.~(\ref{delalphaini})
and (\ref{barX1/n+q}). 
The optimal width $\bar{X}=1/q_n$ 
for the main sequence is given in Table~\ref{tab:2111}.

\begin{figure}
\begin{center}
\includegraphics{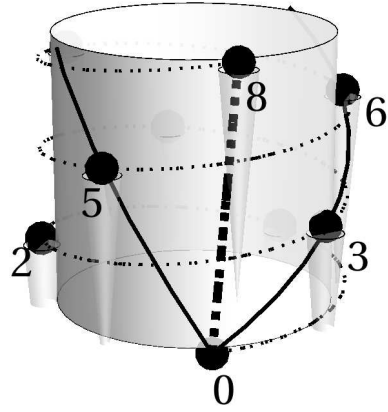}
\caption{\label{parastichy}
A 'vascular leaf-trace primordium' of an 8-parastichy (dashed line) 
for $\alpha=3/8+\delta \alpha$ ($\delta \alpha=0.003, n_c=6$) 
is shown along with a parastichy pair (3,5) (solid lines).
The sign of $\delta \alpha$ coincides with the sign of $2/5-3/8>0$, 
thereby the divergence angle $\alpha$ is slightly shifted
from the PF 3/8 toward a limit $\alpha=0.382$.
}
\end{center}
\end{figure}

Strictly speaking, 
the minimum is not at $\alpha=\bar{\alpha}$ in  Eq.~(\ref{alpha0}),
but lies at $\alpha=\bar{\alpha} + \delta \alpha$ 
by which a small correction $\delta \alpha$ is defined.
Let us find $\delta \alpha$ by the condition $v'(\bar{\alpha}+\delta \alpha)=0$. 
In the lowest order in $\delta \alpha$, 
we have 
\begin{equation}
v'\left(\bar{\alpha}+\delta \alpha\right)=
(qv_q-n v_n    )V' \left(\bar{X}\right)
+ (n^2 v_n  +q^2 v_q) \delta \alpha V'' \left(\bar{X} \right). 
\end{equation}
Hence we find
\begin{equation}
\delta \alpha= 
\frac{n v_n-qv_q    }{ n^2 v_n  +q^2 v_q  }
\left(
\frac{- V' \left(\bar{X}\right)}
{V'' \left(\bar{X} \right)}
\right).
\label{delta alphanvn}
\end{equation}
By the assumption that $V(x)$ is  localized with a half width
$X$ around $x=0$, 
it is physically reasonable to suppose
$V' \left(\bar{X}\right)<0$ and $ V'' \left(\bar{X} \right)>0$ at a tail of $V(x)$. 
Accordingly, 
the sign of $\delta \alpha$ is  determined 
by the difference between the denominators of the parent fractions $m/n$ and $p/q$.
Under a weak condition that 
$v_n$ decreases 
more rapidly than $1/n$,
{\it 
around a mediant $\bar{\alpha} =(m+p)/(n+q)$ between 
parent fractions $m/n$ and $p/q$, 
a minimum of $v(\alpha)$ 
is slightly shifted from $\alpha=\bar{\alpha}$ 
to the side of $m/n$ or $p/q$ with a larger denominator}.
This is consistent with 
Levitov's maximal denominator principle on a Farey tree (\cite{levitov91b}).
%
For instance, the position of the minimum in Fig.~\ref{pot38}
is slightly shifted from 3/8 to the side of 2/5 (instead of 1/3).  
The effect of $\delta \alpha$ is neglected in Sec.~\ref{sec:result}.
It is surely negligible mostly as confirmed explicitly.
Still there are cases where the deviation $\delta \alpha$ has discernible effects. 
First of all, non-zero $\delta \alpha$
deforms an `orthostichy' $\alpha=(m+p)/(n+q)$ into a parastichy (\ref{sec:mediant}). 
According to the above theorem, 
the sign of $\delta \alpha= \alpha- (m+p)/(n+q)$ 
is given by the sign of $m/n- (m+p)/(n+q)$ $(n>q)$ (Fig.~\ref{parastichy}).
{\it  
The deviation from a fraction 
$\delta \alpha$ 
tends toward 
a limit divergence angle} (Sec.~\ref{sec:majorseq}).

%

Dynamical models aim to derive $\alpha=1/\tau^2$ 
deterministically
(\cite{adler74,rk89b,levitov91b,dc96}).
Their results are consistent with each other.
The original idea of the dynamical mechanism can be traced back to
\cite{vaniterson} and \cite{schwendener}.
Most of them are 
based on a {\it geometric} assumption that the interaction $V(|r_m-r_n|)$ depends only on 
the distance $|r_m-r_n|$ between the two points $n$ and $m$.
Generally,  in an anisotropic system like plants, the assumption does
not hold true.
In short, a leaf 
 is not a sphere on a cylinder surface (Fig.~\ref{para25}).
If we had adopted
this strong assumption,
we would have failed to reproduce PFs.
This is because 
$\delta \alpha$ enforced  by the assumption is relatively large and not negligible.
To compare with our model, however, it must be remembered that dynamical models mostly follow a long tradition of reseach focused on shoot apices. Unlike ours, they are not aimed at PFs of mature leaves.

Finally, as a possible generalization, 
let us mention a dynamical variant of our model. 
In the dynamical model, 
any slight deviation  $\delta \alpha$
can be effectively important
to determine a dynamical path of $\alpha$.
In the main text, 
we assumed that $\alpha_{0}$ is a preset constant independent of $n_{c}$.
In contrast, 
one may consider a model in which $\alpha_{0}$ 
is variable 
such that 
$\alpha=\alpha_{\rm min}$ 
to minimize $v(\alpha)$ at a growth step $n_{c}$ determines 
the initial value $\alpha_{0}$ of the next step $n_{c}+1$, i.e., 
\begin{equation}
\alpha_{0}(n_{c}+1)  = \alpha_{\rm min} (n_{c}), 
\end{equation}
or let $\alpha_{0}$
evolve continuously or adaptively
along a branch in the tree of Fig.~\ref{tree}.
To begin with, 
the initial condition must be set for $\alpha_{0} (1)$. 
To avoid confusion, $\alpha_{0}$ of this model should rather be
written as $\alpha$.
Consider what happens when a local maximum 
begins to appear around 
a minimum at $\alpha=\alpha_{\rm min}$, 
when 
$n_{c}$ reaches $n_{0}+\Delta n$ for $\alpha_{\rm min}$.
The maximum occurs just at a rational number near but not at
$\alpha_{\rm min}$. 
As discussed above, 
$\alpha_{\rm min}$ is shifted by a finite amount $\delta \alpha$ from the position of the maximum.
Therefore, 
the sign of $\delta \alpha$ uniquely determines the next minimum to be chosen.
Between two newborn daughter fractions, 
the one with a larger denominator is always chosen
(a daughter fraction  with a larger denominator 
lies on the same side of 
a parent fraction with a larger denominator).  
Thus, 
we obtain $\alpha\gtrsim 1/3$ at $n_c=2$, 
then $\alpha\lesssim 2/5$ at $n_c=3$, 
and  $\alpha\gtrsim 3/8$ at $n_c=5$, then 
$\alpha\lesssim 5/13$ at $n_c=8$,  and so forth.
By following the main branch of Fig.~\ref{tree}, 
we reach $\alpha=1/\tau^2$ finally in the limit
 $n_{c}\rightarrow \infty$. 
The bifurcation rule of 
this dynamical mechanism accords with 
(H) in Sec.~\ref{subsec:main}.
Nonetheless, 
the causal relationship between PFs and the golden angle is reversed here.
In the dynamical mechanism, the golden angle is the {\it effect}, or the end result
$\alpha(\infty)$ to be obtained generally. 
In the static mechanism, it is the {\it cause} $\alpha(0)$ to be selected specifically.

\end{document}